\newcommand{\version}{February 9, 2021}
\documentclass[letterpaper,oneside,11pt,pdftex]{article}
\pdfoutput=1
\usepackage[bindingoffset=0.0cm,textheight=22.60cm,hdivide={*,16.25cm,*}, vdivide={*,22.60cm,*}]{geometry}
\usepackage{amsmath,amsfonts,amssymb}
\usepackage{mathtools}
\usepackage[pdftex]{graphicx}
\usepackage[margin=1cm,font=small]{caption}
\usepackage[T1]{fontenc}
\usepackage[utf8]{inputenc}
\usepackage{fouriernc}

\usepackage[numbers,square,comma,sort&compress]{natbib}
\usepackage{import}

\usepackage[pdftex,hyperref,svgnames]{xcolor}
\usepackage[pdftex,bookmarksnumbered=true,breaklinks=true,%
colorlinks=true,linktocpage=true,linkcolor=MediumBlue,citecolor=ForestGreen,urlcolor=DarkRed]{hyperref}

\makeatletter
\AtBeginDocument{
  \hypersetup{
    pdfkeywords = {\keywords},
    pdftitle = {\@title},
    pdfauthor = {\@author}
  }
}
\makeatother

\makeatletter

 \@addtoreset{equation}{section}
 \makeatother

\renewcommand{\d}{\delta}
\newcommand{\e}{\epsilon}

\newcommand{\vth}{\vartheta}

  \newcommand{\cF}{\mathcal{F}}
  
  \newcommand{\cL}{\mathcal{L}}
  \newcommand{\cO}{\mathcal{O}}
  \newcommand{\cR}{\mathcal{R}}

\newcommand{\sgn}[1]{\textrm{sgn}\!\left(#1\right)}
\newcommand{\nn}{\nonumber}
\newcommand{\eqnref}[1]{Eq. \eqref{#1}}

\newcommand{\ct}{c_{\textrm{T}}}

\newcommand{\gt}{\gamma_{\textrm{T}}}

\newcommand{\bt}{\beta_{\textrm{T}}}

\newcommand*{\mat}[1]{\mathrm{#1}}
\renewcommand{\Im}{\mathrm{Im}}
\renewcommand{\Re}{\mathrm{Re}}

\newcommand{\txt}[1]{\textrm{#1}}

\DeclarePairedDelimiter\abs{\lvert}{\rvert}
\newcommand{\coleq}{\vcentcolon=}

\title{\texorpdfstring{\begin{flushright}
        {\small LA-UR-20-25951}
       \end{flushright}\vspace{2em}}{}%
       A general solution for accelerating screw dislocations in arbitrary slip systems with reflection symmetry}

\author{Daniel N. Blaschke}

\date{\version}

\newcommand{\keywords}{dislocations in crystals, dislocation mobility, crystal plasticity, transsonic motion}

\graphicspath{{./}{./figures/}}

\begin{document}

 \maketitle

 \thispagestyle{empty}
 \begin{center}
 \vspace{-0.3cm}
 Los Alamos National Laboratory, Los Alamos, NM, 87545, USA
 \\[0.5cm]
 \ttfamily{E-mail: dblaschke@lanl.gov}
 \end{center}

\begin{abstract}
Solutions to the differential equations of linear elasticity in the continuum limit in arbitrary crystal symmetry are known only for steady-state dislocations of arbitrary character, i.e. line defects moving at constant velocity.
Troubled by singularities at certain `critical' velocities (typically close to certain sound speeds), these dislocation fields are thought to be too idealized, and divergences are usually attributed to neglecting the finite size of the core and to the restriction to constant velocity.
In the isotropic limit, accelerating pure screw and edge dislocations were studied some time ago.
A generalization to anisotropic crystals has been attempted for pure screw and edge dislocations only for some special cases.
This work aims to fill the gap of deriving a general anisotropic solution for pure screw dislocations applicable to slip systems featuring a reflection symmetry, a prerequisite to studying pure screw dislocations without mixing with edge dislocations.
Further generalizations to arbitrary mixed dislocations as well as regularizations of the dislocation core are beyond the scope of this paper and are left for future work.
\end{abstract}

 \newpage
\tableofcontents

\section{Introduction and background}
\label{sec:intro}

Plasticity in crystalline materials is governed by dislocations
and at high stress and temperature their mobility becomes increasingly important, as it determines the glide time between obstacles (grain boundaries, impurities, other defects, etc.), thereby affecting Orowan's relation \cite{Hansen:2013,Luscher:2016}.
Velocity $v$ depends on dislocation drag $B$ (the impediment of dislocation movement due to interaction with phonons, etc.) in a non-linear manner, and a reliable model of dislocation displacement gradient fields is a prerequisite for any model of $B(v)$ \cite{Nadgornyi:1988,Alshits:1992,Blaschke:2019Bpap,Blaschke:2018anis}, an important ingredient of both discrete dislocation dynamics (DDD) simulations \cite{Zbib:1998,Ghoniem:2000,Bertin:2015,Cui:2019} and plasticity models \cite{Lloyd:2014JMPS,Luscher:2016,Blaschke:2019a}.
Yet dislocation mobility in this regime is poorly understood, posing a major roadblock to further improving material strength predictions at high stress and temperature.
A key question in this regard is whether dislocations can reach transonic and supersonic speeds under sufficiently high stress.
Until some years ago it was believed that they cannot \cite{Weertman:1980}, based on the linear-elasticity derivation that the elastic self-energy of dislocations moving at constant velocity in the isotropic limit diverges at the transverse sound speed.
However, in 2009 supersonic dislocations were observed in a plasma crystal \cite{Nosenko:2007}.
Also, molecular dynamics (MD) simulations of screw and edge dislocations in some metals suggest that dislocation velocities can reach or even exceed the transverse wave speed \cite{Olmsted:2005,Marian:2006,Tsuzuki:2008,Oren:2017,Peng:2019,Blaschke:2020MD}.
Experiments cannot track dislocations in real time at these high speeds, but some crucial information is contained in measured stress versus strain rate curves.
In particular, the highest measured strain rates (produced by overdriven shocks) are consistent with dislocations moving close to the lowest shear wave speed as a lower bound if, and only if, the density of mobile dislocations is comparable to the highest measured total dislocation density \cite{Blaschke:2019a}.
Despite positive MD results, clarifying whether supersonic dislocations exist in real metals will hence not only require a good understanding of the dislocation displacement gradient field itself, but will also require studying the dislocation density evolution; we will only focus on the dislocation field and not attempt the latter here.

Regarding dislocation theory, the displacement field of dislocations is determined by the equations of motion and the (leading order) stress-strain relations.
Their solutions in the isotropic steady-state limit are well-known \cite{Weertman:1980} and suffer from divergences at the transverse and longitudinal sound speed.
See also \cite{Rosakis:2001} for a treatment of the subsonic, transonic, and supersonic regimes in the isotropic limit where each regime is separated from the others by a divergence in the dislocation field.

Solutions for accelerating screw and edge dislocations in the isotropic limit were derived by Markenscoff et al. \cite{Markenscoff:2008,Markenscoff:2009,Huang:2009}, showing that an acceleration term together with a regularized dislocation core removes the divergence, thereby opening the possibility of supersonic events.
An alternative treatment of the isotropic case which also emphasizes the need to account for size variations of the core can be found in a separate series of papers, Refs. \cite{Pillon:2007,Pellegrini:2010,Pellegrini:2014,Pellegrini:2020}.

Less is known in the more general anisotropic case, which is important because dislocation glide motion always happens within anisotropic single crystal grains (even in polycrystals):
Solutions for the dislocation field for arbitrary anisotropic crystals are known in the simple case of constant dislocation velocity \cite{Bacon:1980}.
A general method developed by Stroh and others \cite{Bacon:1980} (and recently refined by Pellegrini \cite{Pellegrini:2017}) can be used to derive solutions which suffer from diverging self-energies at the various wave speeds, suggesting that the lowest of those velocities is an upper bound.
Ref. \cite{Pellegrini:2018} also introduced a method to include an extended dislocation core with elliptic shape into this steady-state framework.
For general (non-constant) dislocation motion, solutions are known only for some special cases regarding crystal symmetry and slip planes.
In particular, Markenscoff and Ni \cite{Markenscoff:1984JE,Markenscoff:1984,Markenscoff:1985a} derived the solution for an accelerating pure screw
and a pure edge dislocation in coordinates aligned with the dislocation and restricted to special slip planes in cubic and hexagonal crystals (such as basal slip).
The pure screw solution of Ref.  \cite{Markenscoff:1984JE} relies on a coordinate transformation bringing the underlying differential equations back into the form of the isotropic case.
Since the boundary condition is not transformed in the same way, the dislocation motion in that work is in a different direction than is usually considered (in contrast to our present study), i.e. part of the motion considered in \cite{Markenscoff:1984JE}\footnote{
I thank one of the anonymous referees for making me aware of this reference.
}
is in the direction that would be the slip plane normal for a mixed dislocation (though in the strictly pure screw case the slip plane is not well defined).
Another special-case solution for moving edge dislocations in crystals with transverse (with respect to the slip plane) isotropy, and hence applications to e.g. basal slip in hcp crystals, (using a different method than Markenscoff) was carried out by Payton \cite{Payton:1985,Payton:1995}.
For a recent nice review of dislocation dynamics, see Ref.~\cite{Gurrutxaga:2020}.

The present work is a step towards further generalizing Markenscoffs results for accelerating dislocations to arbitrary anisotropic crystals.
In particular, we study pure screw dislocations and, as a first step, neglect the dislocation core.
This can only be done in slip systems exhibiting a reflection symmetry \cite[Chapter 13]{Hirth:1982}, since only then do pure screw and edge dislocations decouple.
All 12 fcc slip systems as well as a number of other slip systems in hcp and some crystals with even lower symmetries share this property, though none of the 48 bcc slip systems do.
As was pointed out in Refs. \cite{Teutonico:1961,Blaschke:2020MD}, steady-state pure screw dislocations diverge at a `critical' velocity that is different than any sound speed.
Here, we confirm for anisotropic crystals that (similar to the isotropic limit) this divergence
is still present when acceleration is taken into account.
Regularizing the dislocation core is hence required to eliminate the divergence entirely.

We also emphasize, that our present results represent the most general solution for accelerating pure screw dislocations in anisotropic crystals to date, which are interesting in their own right as well.
In particular, we pay attention to the various slip systems in anisotropic crystals, assume our dislocation glides in the physically relevant directions, and take into account all velocity regimes.
As such, the solution derived here is applicable to all those crystal symmetries and slip systems whose glide planes are reflection planes.
This includes not only all 12 fcc slip systems, but likewise also hcp, tetragonal, orthorhombic, trigonal, and (possibly) other crystal geometries exhibit a number of slip planes which fulfill this symmetry requirement.
In studying various limits, we recover many other known results, such as steady-state screw dislocations for subsonic and supersonic regimes and the isotropic limit.

\section{Solving the eom with Laplace and Fourier transforms}
\label{sec:laplace}

The governing equations to which the displacement gradient field $u_{i,j}\coleq\partial_j u_i$ provides a solution are the equations of motion and the (leading order) stress-strain relations known as Hooke's law:
\begin{align}
 \partial_i\sigma_{ij}&=\rho\ddot u_j\,, &
 \sigma_{ij}&= 
C_{ijkl}\epsilon_{kl}=
C_{ijkl} \partial_l u_{k}
\,, & \nonumber\\
 \epsilon_{kl}&:=(\partial_lu_{k}+\partial_k u_{l})/2
 \,, \label{eq:Hooke}
\end{align}
where $\sigma_{ij}$ denotes stress, $\epsilon_{kl}$ is the infinitesimal strain tensor (i.e. the symmetrized displacement gradient field), and $\rho$ the material density.

For example, assuming cubic symmetry (e.g. fcc or bcc), the tensor of second order elastic constants within the crystal reference frame may be written as
\begin{align}
 C_{ijkl}&=c_{12}\d_{ij}\d_{kl}+c_{44}\left(\d_{ik}\d_{jl}+\d_{il}\d_{jk}\right)-\left(2c_{44}+c_{12}-c_{11}\right)\sum_{\alpha=1}^3\d_{i\alpha}\d_{j\alpha}\d_{k\alpha}\d_{l\alpha}
 \,,\label{eq:C2_cubic}
\end{align}
where the first two terms are invariant under rotations.
The last term explicitly depends on the crystal basis vectors, which in the present (Cartesian) case coincide with $\hat x_i=\d^1_i$, $\hat y_i=\d^2_i$, and $\hat z_i=\d^3_i$, and as such it must be transformed into the coordinate basis one wishes to perform calculations in.

\subsection{The general case}
\label{sec:generalcase}

For some slip system geometries (but not all), pure screw and edge dislocations can be treated separately.
This is only possible if a pure screw component gives rise \emph{only} to displacements $u_z$ in coordinates where $\hat z$ is aligned with the dislocation line sense (and equally an edge component gives rise only to displacements $u_x$ and $u_y$).
As discussed in Ref. \cite[Chapter 13]{Hirth:1982}, this is the case for slip systems where the $x$, $y$ plane (in coordinates aligned with the dislocation) is a reflection plane, since only then does $u_x=0=u_y$ imply the vanishing of stresses in the $x-y$ plane, i.e. $\sigma_{xx}=\sigma_{yy}=\sigma_{xy}=0$.
For example, in cubic crystals this condition is fulfilled for all 12 fcc slip systems, but not for any bcc slip systems.
Additionally, a number of hcp, tetragonal, orthorhombic, and trigonal slip systems also fulfill the symmetry requirements of the present derivation.

The differential equation \eqref{eq:Hooke} was written in terms of Cartesian coordinates aligned with the
crystal axes.
Solving it for a pure screw dislocation is however more conveniently carried out in coordinates aligned with the dislocation.
Hence, we presently choose our coordinate system with $\hat z$ aligned with the dislocation line and $\hat y$ with the slip plane normal.
By definition, the Burgers vector is aligned with the dislocation line sense of a pure screw dislocation, and provided the symmetry requirements described above are fulfilled for a slip system of interest, the displacement vector for pure screw dislocations will take the form $\vec{u}=(0,0,u_z(x,y,t))$.
Assuming the dislocation is much longer than a Burgers vector, the only velocity component that matters for dislocation glide is then normal to the dislocation line and therefore parallel to $\hat{x}$.
In the rotated coordinate frame, the differential equation for a pure screw dislocation reads
\begin{align}
\rho\partial_t^2u_z(x,y,t)=(A\partial_x^2+B\partial_x\partial_y+C\partial_y^2)u_z(x,y,t)
\,, \label{eq:diffeq_screw_gen}
\end{align}
where numerical coefficients $A$, $B$, and $C$ are functions of the second order elastic constants ($c_{11}$, $c_{12}$, and $c_{44}$ in the case of cubic symmetry) as well as the rotation matrix that transforms between the crystal coordinates and our present coordinates which in turn depend on the slip system geometry.
Coefficients $A$, $B$, and $C$ for the fcc slip systems were previously presented in Ref. \cite{Blaschke:2020MD} in the context of the steady-state limit of \eqnref{eq:diffeq_screw_gen}.
For completeness, we derive these coefficients once more in Appendix \ref{sec:rotmat}.

The boundary conditions appropriate for a screw dislocation with Burgers vector $b\hat{z}$ are
\begin{align}
u_z(x,y\to0^\pm,t) &=\pm\frac{b}{2}\Theta(x-l(t))
\,, &
\sigma_{yy}(x,0,t) = 0
\qquad  \qquad \forall t\ge0
\,, \label{eq:bc_iso}
\end{align}
where $\Theta(x)$ denotes the Heaviside step function.
We assume here that the dislocation is initially at rest for $\forall t<0$ and starts to move according to $x=l(t)$ from $t\ge0$.
The second condition encodes the requirement that no external concentrated force need to be applied in the $y$-direction at the core of the dislocation, and it is automatically fulfilled by a pure screw dislocation since $u_y=0$.
The first boundary condition encodes the discontinuity upon crossing the slip plane from negative to positive $y$, i.e. the displacement changes sign when approaching $y\to0$ from above or below the slip plane.

The static case is included by these boundary conditions upon setting $l(t)=0$
and the solution is then straightforwardly checked to be \cite{Blaschke:2020MD}
\begin{align}
u_z^\txt{stat}(x,y,t) 
&= \frac{b}{2\pi} \arctan\left(\frac{y\sqrt{\frac{1}{C} \left(A -  \frac{{B}^2}{4C}\right)}}{-x+\frac{{B}}{2C}y}\right)
\,. \label{eq:uz_static_anis}
\end{align}
In the general case with $l(t)\neq0$, it is convenient to employ a Laplace transform in time $t$ and a Fourier transform (or two-sided Laplace transform which is related to the former) in one spatial variable, and it is convenient to transform $x$ in this case due to the boundary condition \eqref{eq:bc_iso}.
In particular:
\begin{align}
\cL\{u\}&=\int_0^\infty u e^{-st} dt
\,,&
\cF\{u\}&=\frac{1}{\sqrt{2\pi}}\int_{-\infty}^\infty u e^{ikx} dx
\,, \label{eq:bothtransforms}
\end{align}
and it will be convenient to write the Fourier transform of $u$ as a function of $\alpha\coleq k/s$.
Furthermore, we will make use of the Cagniard-de Hoop method \cite{Cagniard:1939,DeHoop:1960,Freund:1973,Markenscoff:1980} whose strategy it is to perform the integration of the \emph{inverse} Fourier transform along such a path that the resulting integral can be recognized as the Laplace transform of a certain function of time\footnote{
The same methods have also been applied to the theory of seismic faults which can be described in terms of isotropic dislocations, see e.g. \cite{Mitra:1966,Boore:1971,Boore:1974,Madariaga:1978} and references therein.
I thank B. Gurrutxaga-Lerma for pointing me to these references.
}.
Applying both transforms \eqref{eq:bothtransforms} to the differential equation \eqref{eq:diffeq_screw_gen} after dividing by $A$, we find
\begin{align}
0&=\frac{1}{\sqrt{2\pi}}\int_{-\infty}^\infty dx \int_0^\infty dt\, e^{-st+is\alpha x}\left[\frac{s^2}{c_A^2} + s^2\alpha^2 + i\tilde{B}s\alpha \partial_y - \tilde{C} \partial_y^2\right]u_z(x,y,t)
\nn\\
&=\left[s^2\left(\alpha^2+\frac{1}{c_A^2}\right) + i\tilde{B}s\alpha \partial_y - \tilde{C}\partial_y^2\right]\cF\{\cL\{u_z\}\}(\alpha,y,s)
\,,\label{eq:diffeqnlaplace}
\end{align}
where $c_A=\sqrt{A/\rho}$, $\tilde{C}=C/A$, and $\tilde{B}=B/A$.
Using the method of split variables and requiring the solution be bounded as $y\to\pm\infty$, we deduce
\begin{align}
\cF\{\cL\{u_z\}\}(\alpha,y,s) = U(\alpha,s) e^{-s\tilde{\beta} y}
\,, \label{eq:ansatz_gen}
\end{align}
with $\sgn{y}\Re(\tilde{\beta})>0$.
We find
\begin{align*}
\tilde{C}\tilde{\beta}^2 + i\tilde{B}\alpha \tilde{\beta} -\left(\alpha^2+\frac{1}{c_A^2}\right) = 0
\,,
\end{align*}
and hence 
\begin{align}
\tilde{\beta}
&=\frac1{2\tilde{C}}\left(-i\tilde{B}\alpha + \sgn{y} \sqrt{4\tilde{C}\left(\alpha^2+\frac{1}{c_A^2}\right)-\tilde{B}^2\alpha^2}\right)
\nn\\
&=\sgn{y}\sqrt{\frac{\alpha^2}{\tilde{C}}\left(1-\frac{\tilde{B}^2}{4\tilde{C}}\right)+\frac{1}{c_A^2\tilde{C}}}-i\frac{\tilde{B}\alpha}{2\tilde{C}}
\,.
\end{align}
Assuming that $\tilde{B}^2<4\tilde{C}$, we must choose the sign of the square root according to $\sgn{y}$ and hence find that\
\begin{align}
\tilde{\beta} y &=\beta\abs{y}
 \,, &
\beta &= \left(\sqrt{\frac{1}{c_A^2\tilde{C}}-\frac{p^2}{\tilde{C}}\left(1-\frac{\tilde{B}^2}{4\tilde{C}}\right)}-\sgn{y}\frac{\tilde{B}p}{2\tilde{C}}\right)
\,,\label{eq:betaofpanis}
\end{align}
with $p=i\alpha=ik/s$ being a further variable substitution whose purpose will become clear below.
In Appendix \ref{sec:rotmat} we check explicitly that the condition $\tilde{B}^2<4\tilde{C}$ is indeed fulfilled for all 12 fcc slip systems, and one may check that this also the case for a number of hcp and other slip systems with reflection symmetry.

Making use of our boundary conditions, we determine $U(\alpha,s)$ as follows:
\begin{align}
U(\alpha,s) &= \cF\{\cL\{u_z(x,0^\pm,t)\}\} = \frac{\pm1}{\sqrt{2\pi}}\int_{-\infty}^\infty dx \int_0^\infty dt\, e^{-st+is\alpha x} \frac{b}{2}\Theta(x-l(t))
\nn\\
&= \pm\frac{b}{2\sqrt{2\pi}}\int_{0}^\infty dx \int_0^{\eta(x)} dt\, e^{-st+is\alpha x} 
\nn\\
&= \frac{\pm b}{2\sqrt{2\pi}}\int_{0}^\infty dx\, e^{is\alpha x} \frac{1}{s}\left(1-e^{-s\eta(x)}\right)
\,,\label{eq:Uas_iso}
\end{align}
where $\eta(x)\coleq {l^{-1}(x)}$ is strictly positive since the integral is over $t>0$, and we consider $l(t)$ to be a monotonically growing (or decreasing) function of $t$.
For non-vanishing $y$, the correct sign is encoded by $\sgn{y}$.
The first term within \eqref{eq:Uas_iso} is recognized as the static solution since $l(t)\to0$ implies $e^{-s\eta(x)}\to0$.

The next step is to apply the inverse Fourier transform, and in doing so we make use of the static solution \eqref{eq:uz_static_anis} for the $\eta$-independent term in order to avoid unphysical singularities in later steps\footnote{
The author thanks X. Markenscoff for clarifying this subtle point.
}, a method previously employed in Ref. \cite{Markenscoff:1980};
we presently have
\begin{align}
\cL\{u\}(x,y,s) &= \frac{1}{\sqrt{2\pi}}\int_{-\infty}^\infty dk e^{-ikx}U(\alpha=k/s,s) e^{-s\beta \abs{y}}
\nn\\
&= \cL\{u_z^\txt{stat}\}(x,y,s)
 - \frac{b\,\sgn{y}}{4\pi}\int_{-\infty}^\infty dk e^{-ikx}\int_{0}^\infty dx'\, e^{ik x'} \frac{1}{s} e^{-s\eta(x')} e^{-s\beta \abs{y}}
\,. \label{eq:Luansatz}
\end{align}
The integral over $x'$ can be solved analytically only for simple special cases of $\eta(x')$, like for example for a dislocation initially at rest and then `suddenly' moving at constant velocity $v$ from time $t>0$ leading to $\eta(x')\to x'/v$ (with $x'$ positive).
We will for now keep $\eta(x')$ general which requires us to exchange the two integrals in the second term and to solve for the integral over $dp$ first there.
This is permissible only if both integrals converge absolutely, but in the expression above this is not the case as discussed in Ref. \cite{Markenscoff:1980} (for the isotropic limit), and as we will also see below.
In fact, solving for $\cL\{u\}(x,y,s) $ directly is troublesome due to subtleties with regard to poles, and instead we proceed to solve for its gradient.
Furthermore, the ansatz we made for \eqref{eq:Luansatz}
(using the method of split variables)
is good for the $y>0$ and $y<0$ half planes separately, but is troublesome at $y=0$ because a distributional term $\partial_y\sgn{y}=2\delta(y)$ in its gradient would violate the differential equation \eqref{eq:diffeqnlaplace} at $y=0$.
Both subtleties are elegantly avoided by solving directly for the gradient of displacement field $u$ in the two half planes separately and then continuously gluing them together at $y=0$.
This is the same strategy previously employed in Refs. \cite{Freund:1973,Markenscoff:1980} in the isotropic case.
We presently have
\begin{align}
\partial_x \cL\{u\}(x,y,s) 
&= \cL\{\partial_x u_z^\txt{stat}\}(x,y,s)  - \frac{ib\,\sgn{y}}{4\pi}\int\limits_{-i\infty}^{i\infty} dp e^{-spx}\int\limits_{0}^\infty dx'\, e^{spx'} sp e^{-s\eta(x')} e^{-s\beta \abs{y}}
\,, \nn\\
\partial_y \cL\{u\}(x,y,s) 
&= \cL\{\partial_y u_z^\txt{stat}\}(x,y,s)
 - \frac{ib}{4\pi} \int\limits_{-i\infty}^{i\infty} dp e^{-spx}\int\limits_{0}^\infty dx'\, e^{spx'} s\beta e^{-s\eta(x')} e^{-s\beta \abs{y}}
\,,
\end{align}
where the variable substitution $k=-ips$ and $dk=-isdp$ was employed
and the sign function allows us to treat both half planes simultaneously\footnote{
We subsequently see from \eqnref{eq:accanis} below that the solutions for the two half planes can indeed be continuously glued together everywhere for subsonic dislocations. For supersonic dislocations the same is true everywhere except for the one point corresponding to the position of the (unregularized) dislocation core where the solution is singular and breaks down.
}.
Following the Cagniard-de Hoop method \cite{Cagniard:1939,DeHoop:1960,Markenscoff:1980}, we wish to find a further variable substitution that allows us to rewrite the integral over the purely imaginary variable $p$ in terms of a strictly positive variable $\tau$ such that the integrand in terms of $\tau$ is recognized as a Laplace transform of a certain function of time.
This can be done via complex analysis by identifying an appropriate closed path in complex space over which to integrate and by using Cauchy's theorem stating that the integral over such a closed path equals the residua of any poles enclosed by that path.

In the isotropic limit, it was shown in Ref. \cite{Markenscoff:1980}, that after exchanging the integrals, the remaining integral over $x'$ exhibits a quadratic pole in its integrand at $x'\to x$ if $y=0$ and $x\ge0$ and converges otherwise.
We will see below in \eqnref{eq:cFpoles} that this is also the case in the present anisotropic generalization, and in order to remove this (and a subleading) pole we add and subtract a term with $\eta(x')$ replaced by its linear order Taylor series expansion\footnote{Note that the term to be added and subtracted in Ref. \cite[Eq. (24)]{Markenscoff:1980} has the wrong dimensions and is clearly missing a factor $\beta e^{(-s\lambda\xi)}$ where the notation in that paper corresponds to ours via $\lambda=p$ and $\xi=x'$; otherwise the general strategy of removing the pole is the same as ours.
We also point out that the importance of adding and subtracting such terms was not emphasized in some later papers by the same author \cite{Markenscoff:2008,Markenscoff:2009},
and this may lead to the impression that the isotropic solution is highly singular at $y=0$; see e.g. an according comment in \cite{Pellegrini:2010}.
In fact, as we see in our present derivation, the solution is well-behaved if appropriate terms are added and subtracted for all values of $y$ prior to exchanging the order of integration.}
$\tilde\eta=\eta(x)+(x'-x)\eta'(x)$ where we define $\eta'(x)\coleq\sgn{x}\partial_x\eta(\abs{x})\ge0$ and $\eta(x)\coleq\sgn{x}\eta(\abs{x})$.
By changing the overall sign of $\tilde\eta$ according to the sign of $x$ (and we have this freedom as there is no pole for $x<0$), we ensure that the two $x'$ integrals cancel one another in the special case of constant velocity (see Section \ref{sec:constvel} below), thereby significantly simplifying that special case.

Because of its linear dependence on $x'$ in the exponent, the $x'$ integral can be performed exactly and without exchanging the integral order in the added term.
In the subtracted term, the integral order is exchanged in order to remove the pole.
In particular, we consider
\begin{subequations}
\begin{align}
\partial_x \cL\{u\}(x,y,s) 
&= \cL\{\partial_x u_z^\txt{stat}\}(x,y,s)  - \frac{ib\,\sgn{y}}{4\pi}\left[
 \int\limits_{-i\infty}^{i\infty} dp e^{-spx}\int\limits_{0}^\infty dx'\, e^{spx'} sp e^{-s\tilde\eta} e^{-s\beta \abs{y}}
 \right.\nn\\&\quad\qquad \left.
 + \int\limits_{0}^\infty dx'\left(e^{-s\eta(x')} -e^{-s\tilde\eta} \right)\int\limits_{-i\infty}^{i\infty} dp e^{-sp(x-x')} e^{-s\beta \abs{y}}sp\right]
 \nn\\
 &= \cL\{\partial_x u_z^\txt{stat}\}(x,y,s)  - \frac{ib\,\sgn{y}}{4\pi}\left[
 e^{-s\left(\eta(x)-x\eta'(x)\right)} \int\limits_{-i\infty}^{i\infty} dp e^{-spx}  e^{-s\beta \abs{y}} \frac{p}{\eta'(x)-p}
 \right.\nn\\&\quad\qquad \left.
 + \int\limits_{0}^\infty dx'\left(e^{-s\eta(x')} -e^{-s\tilde\eta} \right)s\int\limits_{-i\infty}^{i\infty} dp e^{-sp(x-x')} e^{-s\beta \abs{y}}p\right]
\,,
\end{align}
\begin{align}
\partial_y \cL\{u\}(x,y,s) 
 &= \cL\{\partial_y u_z^\txt{stat}\}(x,y,s) 
 - \frac{ib}{4\pi}\left[
 e^{-s\left(\eta(x)-x\eta'(x)\right)} \int\limits_{-i\infty}^{i\infty} dp e^{-spx} e^{-s\beta \abs{y}} \frac{\beta}{\eta'(x)-p}
  \right.\nn\\&\quad\qquad \left.
 + \int\limits_{0}^\infty dx'\left(e^{-s\eta(x')} -e^{-s\tilde\eta} \right)s \int\limits_{-i\infty}^{i\infty} dp e^{-sp(x-x')} e^{-s\beta \abs{y}} \beta\right]
 \,,
\end{align}
\end{subequations}
where exchanging the order of integration is now allowed after having subtracted the poles under the $x'$ integral.

\begin{figure}[ht]
\centering
\includegraphics[width=0.41\textwidth]{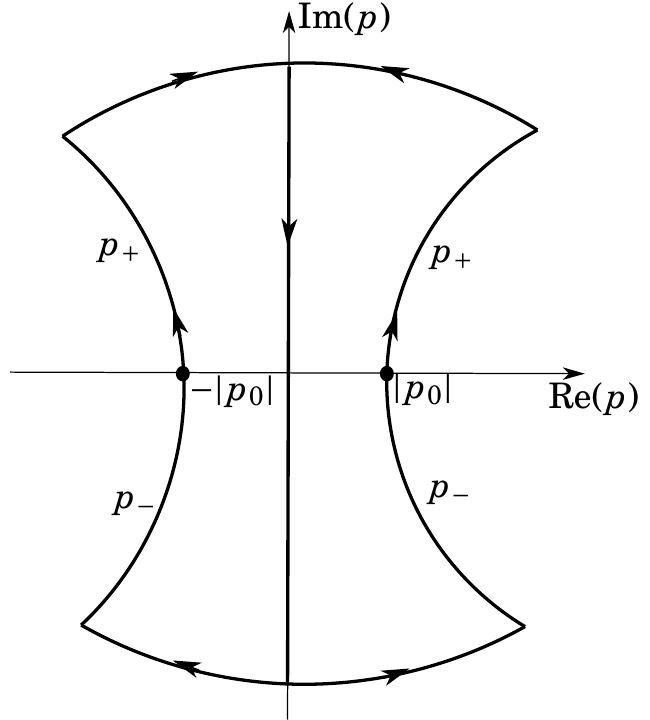}
\caption{We illustrate the path of integration in the complex $p$-plane:
The initial integral over the imaginary axis of $p$ is closed via $p_\pm(\tau)$ which are connected to the imaginary axis at $\Im(p)=\pm\infty$.
The real axis is crossed at $p_0=p(\tau_\txt{min})$ and its sign is given by the sign of $x-y\frac{\tilde{B}}{2\tilde{C}}$.}
\label{fig:thepath}
\end{figure}

\paragraph{Choosing the path:}\ \\
The final step is to choose appropriate paths for the integrals over $p$, so that the integrals are rewritten in a way that can be identified as Laplace transforms in time.
In other words, we need
\begin{align}
(\beta \abs{y}+px)&\equiv \tau>0\,, \qquad \tau\in \Re
\,,\nn\\
(\beta \abs{y}+p(x-x'))&\equiv \tau'>0\,, \qquad \tau'\in \Re
\,,
\end{align}
and $\tau$ (resp. $\tau'$) are subsequently interpreted as a time variable within a Laplace transform.
We proceed by deriving $p(\tau)$ and note that $p(\tau')$ follows trivially by shifting $x\to (x-x')$.
The definition above together with \eqref{eq:betaofpanis} leads to a quadratic equation in $p$, namely
\begin{align}
p^2\left(x^2  - x{y} \frac{\tilde{B}}{\tilde{C}}   + \frac{y^2}{\tilde{C}} \right) - p\tau\left(2x - {y}\frac{\tilde{B}}{\tilde{C}}\right)  + \tau^2- \frac{y^2}{c_A^2\tilde{C}} &= 0
\,,
\end{align}
with solutions
\begin{align}
p_\pm
&= \frac{1}{R^2}\left(\tau\left(x - {y}\frac{\tilde{B}}{2\tilde{C}}\right) \pm i\abs{y}\sqrt{\tau^2\frac{1}{\tilde{C}} \left(1 -  \frac{\tilde{B}^2}{4\tilde{C}}  \right) - \frac{R^2}{c_A^2\tilde{C}} }\right)
\,,\label{eq:ppmanis}
\end{align}
where $R^2 = \left(x^2  - x{y} \frac{\tilde{B}}{\tilde{C}}   + \frac{y^2}{\tilde{C}} \right)$.
The square root within $p_\pm$ is real for $\tau>\tau_\txt{min}=R/\left(c_A\sqrt{1-\tilde{B}^2/4\tilde{C}}\right)\ge0$ since we already established that $1-\tilde{B}^2/4\tilde{C}>0$.
From the expression \eqref{eq:ppmanis}, we see that $\tau\in[\tau_\txt{min},\infty)$ yields hyperbolas in the complex plane of $p_\pm$, as illustrated in Figure \ref{fig:thepath}.
The real axis is crossed at $p_0\coleq p_\pm(\tau_\txt{min})$ with
\begin{align}
\abs{p_0}=\abs{p_\pm(\tau_\txt{min})} &= \frac{\abs*{x - {y}\frac{\tilde{B}}{2\tilde{C}}}}{Rc_A\sqrt{1-\frac{\tilde{B}^2}{4\tilde{C}}}} 
=\frac{\sqrt{R^2 -\frac{y^2}{\tilde{C}}\left(1-\frac{\tilde{B}^2}{4\tilde{C}}\right)}}{R\left(c_A\sqrt{1-\frac{\tilde{B}^2}{4\tilde{C}}}\right)}
\le \frac{1}{\left(c_A\sqrt{1-\frac{\tilde{B}^2}{4\tilde{C}}}\right)}
=\frac{1}{v_\txt{crit}}
\,,\label{eq:thecrossing}
\end{align}
which incidentally is one over the critical velocity in the steady-state limit \cite{Blaschke:2020MD}; we will come back to this point when we study limits of the more general solution we are presently deriving.

The integral over the entire imaginary axis of $p$ becomes a difference of integrals over the positive imaginary axes of $p_\pm$, and we close the path by adding the integrals over $\tau\in[\tau_\txt{min},\infty)$ and by connecting them at $p=\pm i\infty$ where the integrand is zero due to
\begin{align}
\lim_{p\to\pm i\infty}e^{-s\beta\abs{y}}\approx \lim_{p\to\pm i\infty}\exp\left(-s\abs{y}\sqrt{-p^2}\sqrt{\frac{1}{\tilde{C}}\left(1-\frac{\tilde{B}^2}{4\tilde{C}}\right)}-s{y}\frac{\tilde{B}p}{2\tilde{C}}\right)\to 0
\,,
\end{align}
since $(-p^2)\to +\infty$ and the second term in the argument of the exponential is purely imaginary and hence bounded by 1.
By Cauchy's theorem, the integral over the closed path is given by the sum of residua within that path.
In the present case we have a pole at $p=\eta'(x)$ which only needs to be taken into account if $1\le {p_0}/\eta'(x)$ as noted above in \eqref{eq:thecrossing}.
We can ensure this by multiplying the corresponding residuum by an appropriate step function.
The two residuum terms we need are thus straightforwardly computed to be:
\begin{subequations}
\begin{align}
\cR_x &= 2\pi i \txt{Res}\left[\frac{-ib\,\sgn{y}}{4\pi}e^{-sp x} e^{-s\beta \abs{y}}\frac{p}{\eta'(x)-p} \right]
\nn\\
&=\frac{b\,\sgn{y}}{2 }\eta'(x)e^{-s x\eta'(x)} \exp\left[{-s\left( \abs{y}\sqrt{\frac{1}{c_A^2\tilde{C}}-\frac{(\eta'(x))^2}{\tilde{C}}\left(1-\frac{\tilde{B}^2}{4\tilde{C}}\right)}-{y}\frac{\tilde{B}\eta'(x)}{2\tilde{C}}\right)}\right]
\,,
\end{align}
\begin{align}
\cR_y &= 2\pi i \txt{Res}\left[\frac{-ib}{4\pi}e^{-sp x} e^{-s\beta \abs{y}}\frac{\beta}{\eta'(x)-p} \right]
\nn\\
&=\frac{b}{2}\left( \sqrt{\frac{1}{c_A^2\tilde{C}}-\frac{(\eta'(x))^2}{\tilde{C}}\left(1-\frac{\tilde{B}^2}{4\tilde{C}}\right)}-\sgn{y}\frac{\tilde{B}\eta'(x)}{2\tilde{C}}\right)e^{-s x\eta'(x)} 
\nn\\&\quad\times
\exp\left[{-s\left( \abs{y}\sqrt{\frac{1}{c_A^2\tilde{C}}-\frac{(\eta'(x))^2}{\tilde{C}}\left(1-\frac{\tilde{B}^2}{4\tilde{C}}\right)}-{y}\frac{\tilde{B}\eta'(x)}{2\tilde{C}}\right)}\right]
\,.
\end{align}
\end{subequations}
We see that the square root inside the exponentials of $\cR_x$ and $\cR_y$ becomes imaginary for $\eta'(x) > 1/\left(c_A\sqrt{1-\frac{\tilde{B}^2}{4\tilde{C}}}\right)$.
%
For $x - {y}\frac{\tilde{B}}{2\tilde{C}} > 0 $, the closed path is counter clockwise, meaning these residua must be subtracted from the integral over $\tau$ in order to equal the integral over the imaginary axis of $p$.
Otherwise, the path is clockwise and the residua need to be added, in other words we add the residua multiplied by $\sgn{{y}\frac{\tilde{B}}{2\tilde{C}} - x}$.

In rewriting the integrand as a function of $\tau$ we will also need $\beta(\tau)$ and $dp(\tau)$.
From $\tau=(\beta \abs{y}+px)$ we deduce
\begin{align}
\beta_\pm &=\frac{1}{\abs{y}}\left( \tau - p_\pm x\right)
=\frac{1}{R^2}\left[ \tau\left(  \frac{\abs{y}}{\tilde{C}}  - x\frac{\sgn{y}\tilde{B}}{2\tilde{C}}\right) \mp ix\sqrt{\tau^2\frac{1}{\tilde{C}} \left(1 -  \frac{\tilde{B}^2}{4\tilde{C}}  \right) - \frac{R^2}{c_A^2\tilde{C}} }\right]
\,,\label{eq:betapmanis}
\end{align}
and
\begin{align}
dp_\pm &= \frac1{R^2}\left(x - {y}\frac{\tilde{B}}{2\tilde{C}} \pm i\abs{y}\frac{\frac{\tau}{\tilde{C}}\left(1 -  \frac{\tilde{B}^2}{4\tilde{C}}  \right)}{\sqrt{\tau^2\frac{1}{\tilde{C}} \left(1 -  \frac{\tilde{B}^2}{4\tilde{C}}  \right) - \frac{R^2}{c_A^2\tilde{C}}}}\right)d\tau
\,.\label{eq:dppmanis}
\end{align}
All integrals over $p$ now take the general form
\begin{align}
\int\limits_{\tau_\txt{min}}^\infty\!\! dp_+(\tau) f(\tau) - \int\limits_{\tau_\txt{min}}^\infty\!\! dp_-(\tau) f^*(\tau)
+ \sgn{{y}\tfrac{\tilde{B}}{2\tilde{C}} - x}\cR
 = {2i}\Im\left(\,\,\int\limits_{\tau_\txt{min}}^\infty\!\! dp_+(\tau) f(\tau)\right) + \sgn{{y}\tfrac{\tilde{B}}{2\tilde{C}} - x}\cR
\,.
\end{align}
In particular,
\begin{subequations}
\begin{align}
\partial_x \cL\{u\}(x,y,s) 
 &= \cL\{\partial_x u_z^\txt{stat}\}(x,y,s)  + \frac{b\,\sgn{y}}{2\pi}\left[
 e^{-s\left(\eta(x)-x\eta'(x)\right)} \Im\left(\int_{\tau_\txt{min}}^{\infty} dp_+(\tau) e^{-s\tau} \frac{p_+\eta'(x)-\abs{p}^2}{\abs{\eta'(x)-p}^2}\right)
 \right.\nn\\&\quad\qquad \left.
 + \int_{0}^\infty dx'\left(e^{-s\eta(x')} -e^{-s\tilde\eta} \right)s\,\Im\left(\int_{\tau'_\txt{min}}^{\infty} dp'_+(\tau') e^{-s\tau'}p'_+\right)\right]
 \nn\\&\quad
 +\sgn{{y}\frac{\tilde{B}}{2\tilde{C}} - x} \Theta\left(\frac{p_0}{\eta'(x)}-1\right)\cR_x
\,,
\end{align}
\begin{align}
\partial_y \cL\{u\}(x,y,s) 
 &= \cL\{\partial_y u_z^\txt{stat}\}(x,y,s) 
 - \frac{b}{2\pi}\left[
 e^{-s\left(\eta(x)-x\eta'(x)\right)} \Im\left(\int_{\tau_\txt{min}}^{\infty} dp_+(\tau) e^{-s\tau} \frac{\beta_+\left(\eta'(x)-p_-\right)}{\abs{\eta'(x)-p}^2}\right)
  \right.\nn\\&\quad\qquad \left.
 + \int_{0}^\infty dx'\left(e^{-s\eta(x')} -e^{-s\tilde\eta} \right)s\,\Im\left(\int_{\tau'_\txt{min}}^{\infty} dp'_+(\tau') e^{-s\tau'}\beta'_+\right)\right]
 \nn\\&\quad
 +\sgn{{y}\frac{\tilde{B}}{2\tilde{C}} - x} \Theta\left(\frac{p_0}{\eta'(x)}-1\right)\cR_y
 \,,
\end{align}
\end{subequations}
where the primed quantities $p'_\pm$, $\beta'_\pm$, and $\tau'_\txt{min}$ differ from their unprimed counterparts only by $x\to (x-x')$.
Computing the gradient of \eqnref{eq:uz_static_anis} and using the expressions listed in Appendix \ref{sec:buildingblocks} for $\Im(dp_+p_+)$, $\Im(dp_+\beta_+p_-)$, $\Im(dp_+\beta_+)$, $\abs{\eta'(x)-p}^2$, and $\abs{p}^2$ in terms of $\tau$, we find
\begin{subequations}
\begin{align}
&\partial_x \cL\{u\}(x,y,s) 
 =\frac{by}{2\pi R^2}\sqrt{\frac{1}{\tilde{C}} \left(1 -  \frac{\tilde{B}^2}{4\tilde{C}}  \right)}\int\limits_{0}^{\infty}d\tau\, e^{-s\tau} 
 +\sgn{{y}\frac{\tilde{B}}{2\tilde{C}} - x} \Theta\left(\frac{p_0}{\eta'(x)}-1\right) \cR_x
  \nn\\ &
  +\frac{by}{2\pi}\int\limits_{0}^{\infty}\!d\tau\, e^{-s\tau}
 \frac{\Theta\left(\tau\!-\!\tau_\txt{min}\right)\,e^{-s\left(\eta(x)-x\eta'(x)\right)} }{R^2\sqrt{\frac{1}{\tilde{C}} \left(1 -  \frac{\tilde{B}^2}{4\tilde{C}}  \right) - \frac{R^2}{\tau^2c_A^2\tilde{C}}}} \frac{\left[\tau \eta'(x)\left(x - {y}\frac{\tilde{B}}{2\tilde{C}} \right)\frac{1}{\tilde{C}}\left[2\left(1 -  \frac{\tilde{B}^2}{4\tilde{C}}  \right)  - \frac{R^2}{\tau^2 c_A^2} 
 \right] 
 - \left(\tau^2  - \frac{y^2}{c_A^2\tilde{C}} \right)\frac{1}{\tilde{C}}\left(1 -  \frac{\tilde{B}^2}{4\tilde{C}}  \right)\right]}{\left(\tau^2  - \frac{y^2}{c_A^2\tilde{C}} - {2\tau}\eta'(x)\left( x - {y}\frac{\tilde{B}}{2\tilde{C}} \right) + (\eta'(x))^2{R^2}\right)}
 \nn\\
 & + \frac{b}{2\pi}s\int\limits_{0}^{\infty} d\tau\, e^{-s\tau}\int\limits_{0}^\infty dx' 
 \Theta\left(\tau\!-\!\tau'_{\!\txt{min}}\right)\left(e^{-s\eta(x')} -e^{-s\tilde\eta} \right)
\frac{\tau y\left(x-x' - {y}\frac{\tilde{B}}{2\tilde{C}} \right)\left(2\left(1 -  \frac{\tilde{B}^2}{4\tilde{C}}  \right)  - \frac{R'^2}{\tau^2 c_A^2} \right)}{\tilde{C} R'^4\sqrt{\frac{1}{\tilde{C}} \left(1 -  \frac{\tilde{B}^2}{4\tilde{C}}  \right) - \frac{R'^2}{\tau^2 c_A^2\tilde{C}}}}
\,,
\end{align}
\begin{align}
&\partial_y \cL\{u\}(x,y,s) 
=\frac{-bx}{2\pi R^2}
\sqrt{\frac{1}{\tilde{C}} \left(1 -  \frac{\tilde{B}^2}{4\tilde{C}}  \right)} \int\limits_{0}^{\infty} d\tau\, e^{-s\tau}
 +\frac{b}{2\pi}\int\limits_{0}^{\infty} d\tau\, e^{-s\tau}
 \frac{\Theta\left(\tau\!-\!\tau_{\!\txt{min}}\right)}{R^2\sqrt{\frac{1}{\tilde{C}} \left(1 -  \frac{\tilde{B}^2}{4\tilde{C}}  \right) - \frac{R^2}{\tau^2c_A^2\tilde{C}}}}
 \nn\\ &\quad\qquad 
  \times  \frac{e^{-s\left(\eta(x)-x\eta'(x)\right)}}{\left(\tau^2  - \frac{y^2}{c_A^2\tilde{C}} - {2\tau}\eta'(x)\left( x - {y}\frac{\tilde{B}}{2\tilde{C}} \right) + (\eta'(x))^2{R^2}\right)}\left\{\frac{\tau \eta'(x)}{\tilde{C}}\left[ \left(  \frac{y^2}{\tilde{C}}  -x^2\right)\left(1 -  \frac{\tilde{B}^2}{4\tilde{C}}  \right) + x\left(x - {y}\frac{\tilde{B}}{2\tilde{C}} \right)\frac{R^2}{\tau^2c_A^2} \right] \right.
\nn\\&\qquad\qquad\qquad \left.  
  - \frac{1}{\tilde{C}}
  \left[\frac{R^2}{c_A^2}\left(x - {y}\frac{\tilde{B}}{2\tilde{C}} \right) - {x}\left(\tau^2  - \frac{y^2}{c_A^2\tilde{C}} \right)\left(1 -  \frac{\tilde{B}^2}{4\tilde{C}}  \right)\right]\right\}
 \nn\\
&\quad + \frac{b}{2\pi}s\int\limits_{0}^{\infty} d\tau e^{-s\tau}\int_{0}^\infty dx' \, 
\Theta\left(\tau\!-\!\tau'_{\!\txt{min}}\right)
\left(e^{-s\eta(x')} -e^{-s\tilde\eta} \right)
\frac{\tau}{R'^4\tilde{C}\sqrt{\frac{1}{\tilde{C}} \left(1 -  \frac{\tilde{B}^2}{4\tilde{C}}  \right) - \frac{R'^2}{\tau^2c_A^2\tilde{C}}}}
\nn\\
&\qquad\qquad\times
\left[ \left(  \frac{y^2}{\tilde{C}}  -(x-x')^2\right)\left(1 -  \frac{\tilde{B}^2}{4\tilde{C}}  \right) + (x-x')\left(x-x' - {y}\frac{\tilde{B}}{2\tilde{C}} \right)\frac{R'^2}{\tau^2c_A^2} \right]
\nn\\
&\quad +\sgn{{y}\frac{\tilde{B}}{2\tilde{C}} - x} \Theta\left(\frac{p_0}{\eta'(x)}-1\right) \cR_y
\,,
\end{align}
\end{subequations}
with $\tau_\txt{min}={R}/\left({c_A\sqrt{1-\frac{\tilde{B}^2}{4\tilde{C}}}}\right)$, $\tau'_\txt{min}={R'}/\left({c_A\sqrt{1-\frac{\tilde{B}^2}{4\tilde{C}}}}\right)$, and $R'=R\big|_{x\to(x-x')}$.
The inverse Laplace transform can almost be read off from the expressions above, considering the following properties of Laplace transforms  \cite{Kuhfittig:1978}:
\begin{itemize}
\itemsep=0pt
\item multiplication by $e^{-sT}$ corresponds to a translation in time by $T$ and
\item multiplication by $s$ corresponds to a time derivative (plus a boundary term which is zero here due to the step function).
\end{itemize}
%
Putting all the pieces together
our general solution in the anisotropic case is
\begin{subequations}\label{eq:accanis}
\begin{align}
&\partial_x u_z(x,y,t)
=\frac{by}{2\pi R^2}
\sqrt{\frac{1}{\tilde{C}} \left(1 -  \frac{\tilde{B}^2}{4\tilde{C}}  \right)}
+\frac{b}{2\pi}\partial_t\int\limits_{0}^\infty dx' 
\left(\cF_x\left[\eta(x'),t\right] - \cF_x\left[\eta(x) + (x'-x)\eta'(x),t\right]\right)
\nn\\
&\quad +\frac{b}{2\pi}
\frac{y}{R^2}
\left[ \Theta\left(\tau-\tfrac{R}{c_A\sqrt{1-\frac{\tilde{B}^2}{4\tilde{C}}}}\right) 
 \frac{\left[\tau \eta'(x)\left(x - {y}\frac{\tilde{B}}{2\tilde{C}} \right)\frac{1}{\tilde{C}}\left[2\left(1 -  \frac{\tilde{B}^2}{4\tilde{C}}  \right)  - \frac{R^2}{\tau^2 c_A^2} 
 \right] 
 - \left(\tau^2  - \frac{y^2}{c_A^2\tilde{C}} \right)\frac{1}{\tilde{C}}\left(1 -  \frac{\tilde{B}^2}{4\tilde{C}}  \right)\right]}{\sqrt{\frac{1}{\tilde{C}} \left(1 -  \frac{\tilde{B}^2}{4\tilde{C}}  \right) - \frac{R^2}{\tau^2c_A^2\tilde{C}}}\left(\tau^2  - \frac{y^2}{c_A^2\tilde{C}} - {2\tau}\eta'(x)\left( x - {y}\frac{\tilde{B}}{2\tilde{C}} \right) + (\eta'(x))^2{R^2}\right)}\right]
\nn\\
&\quad + \frac{b\,\sgn{y}}{2 }\sgn{{y}\tfrac{\tilde{B}}{2\tilde{C}} - x}\eta'(x)\,\Theta\left(\frac{p_0}{\eta'(x)}-1\right)
\delta\!\left(t - {x}\eta'(x) - \left(\! \abs{y}\sqrt{\frac{1}{c_A^2\tilde{C}}-\frac{(\eta'(x))^2}{\tilde{C}}\left(1-\frac{\tilde{B}^2}{4\tilde{C}}\right)}-{y}\frac{\tilde{B}\eta'(x)}{2\tilde{C}}\!\right)\!\right)
\!,\label{eq:accanis_a}
\end{align}
\begin{align}
&\partial_y u_z(x,y,t)
=\frac{-bx}{2\pi R^2}\sqrt{\frac{1}{\tilde{C}} \left(1 -  \frac{\tilde{B}^2}{4\tilde{C}}  \right)}
+\frac{b}{2\pi}\partial_t\int\limits_{0}^\infty dx' 
\left(\cF_y\left[\eta(x'),t\right] - \cF_y\left[\eta(x) + (x'-x)\eta'(x),t\right]\right)
\nn\\&\quad
+\frac{b}{2\pi}
\frac{1}{R^2} \Theta\left(\tau-\tfrac{R}{c_A\sqrt{1-\frac{\tilde{B}^2}{4\tilde{C}}}}\right) 
 \frac{{\tau \eta'(x)}\left(  \frac{y^2}{\tilde{C}}  -x^2\right)\left(1 -  \frac{\tilde{B}^2}{4\tilde{C}}  \right) + \left(\frac{x\eta'(x)}{\tau}-1\right) 
  \frac{R^2}{c_A^2}\left(x - {y}\frac{\tilde{B}}{2\tilde{C}} \right) + {x}\left(\tau^2  - \frac{y^2}{c_A^2\tilde{C}} \right)\left(1 -  \frac{\tilde{B}^2}{4\tilde{C}}  \right)}{\tilde{C}\sqrt{\frac{1}{\tilde{C}} \left(1 -  \frac{\tilde{B}^2}{4\tilde{C}}  \right) - \frac{R^2}{\tau^2c_A^2\tilde{C}}}\left(\tau^2  - \frac{y^2}{c_A^2\tilde{C}} - {2\tau}\eta'(x)\left( x - {y}\frac{\tilde{B}}{2\tilde{C}} \right) + (\eta'(x))^2{R^2}\right)}
 \nn\\&\quad
 + \frac{b}{2}\sgn{{y}\frac{\tilde{B}}{2\tilde{C}} - x}\left(\sqrt{\frac{1}{c_A^2\tilde{C}}-\frac{(\eta'(x))^2}{\tilde{C}}\left(1-\frac{\tilde{B}^2}{4\tilde{C}}\right)}-\sgn{y}\frac{\tilde{B}\eta'(x)}{2\tilde{C}}\right)\Theta\left(\frac{p_0}{\eta'(x)}-1\right)
\nn\\&\qquad\qquad\times
\delta\left(t - {x}\eta'(x) - \left( \abs{y}\sqrt{\frac{1}{c_A^2\tilde{C}}-\frac{(\eta'(x))^2}{\tilde{C}}\left(1-\frac{\tilde{B}^2}{4\tilde{C}}\right)}-{y}\frac{\tilde{B}\eta'(x)}{2\tilde{C}}\right)\right)
,\label{eq:accanis_b}
\end{align}
\end{subequations}
with
\begin{subequations}\label{eq:cFpoles}
\begin{align}
\cF_x\left[\eta,t\right]
&=\Theta\left(t-\eta-\frac{R'}{c_A\sqrt{1-\frac{\tilde{B}^2}{4\tilde{C}}}}\right)
\frac{ y\left(x-x' - {y}\frac{\tilde{B}}{2\tilde{C}} \right)\left(\frac{2}{\tilde{C} }\left(1 -  \frac{\tilde{B}^2}{4\tilde{C}}  \right)\left(t-\eta\right)^2- \frac{R'^2}{ c_A^2\tilde{C}}\right)}{R'^4\sqrt{\frac{1}{\tilde{C}} \left(1 -  \frac{\tilde{B}^2}{4\tilde{C}}  \right)\left(t-\eta\right)^2 - \frac{R'^2}{ c_A^2\tilde{C}}}}
,
\end{align}
\begin{align}
\cF_y\left[\eta,t\right]
&=\Theta\left(t-\eta-\frac{R'}{c_A\sqrt{1-\frac{\tilde{B}^2}{4\tilde{C}}}}\right)
\frac{\left(\frac{y^2}{\tilde{C}}-(x-x')^2\right)\frac{1}{\tilde{C}}\left(1 -  \frac{\tilde{B}^2}{4\tilde{C}}  \right)\left(t-\eta\right)^2 + (x-x')\left(x-x'-y\frac{\tilde{B}}{2\tilde{C}}\right)\frac{R'^2}{c_A^2\tilde{C}}}{R'^4\sqrt{\frac{1}{\tilde{C}} \left(1 -  \frac{\tilde{B}^2}{4\tilde{C}}  \right)\left(t-\eta\right)^2 - \frac{R'^2}{c_A^2\tilde{C}}}}
\,,
\end{align}
\end{subequations}
and
\begin{align}
R^2 &= \left(x^2  - x{y} \frac{\tilde{B}}{\tilde{C}}   + \frac{y^2}{\tilde{C}} \right)
\,,&
R'^2 &= \left((x-x')^2  - (x-x'){y} \frac{\tilde{B}}{\tilde{C}}   + \frac{y^2}{\tilde{C}} \right)
\,,&
\tau &= t - \left(\eta(x)-x\eta'(x)\right) 
\,,\nn\\
p_0 &= \frac{{x - {y}\frac{\tilde{B}}{2\tilde{C}}}}{Rc_A\sqrt{1-\frac{\tilde{B}^2}{4\tilde{C}}}} 
= \frac{\partial_x R}{v_\text{crit}}
\,.
\end{align}
As noted earlier, $\cF_x(\eta)$ and $\cF_y(\eta)$ exhibit quadratic divergences at $y\to0$ and $x'\to x$ (for $y\neq0$, $R'$ never vanishes for real $x'$ since $\tilde{B}^2/(4\tilde{C})<1$). These poles are subtracted by $\cF_x(\tilde\eta)$ and $\cF_y(\tilde\eta)$:
Since $\tilde\eta$ is the linear order Taylor expansion of $\eta$ around $x'=x$, the linear order Taylor expansions of $\cF_x(\eta)$ and $\cF_x(\tilde\eta)$ (and likewise $\cF_y(\eta)$ and $\cF_y(\tilde\eta)$) are equal to one another, thereby cancelling the leading quadratic and subleading linear poles.
This leaves at most only an integrable logarithmic pole and hence the terms integrated over $x'$ are rendered finite.
Notice that the derivative with respect to time must be performed after the integration over $x'$:
exchanging the order would result in a divergent $x'$ integral due to the square root in the denominator of $\cF$ which is zero when the argument of the step function is zero, i.e. prior to taking the time derivative we have an integrable pole at one edge of the integration domain --- see also Refs. \cite{Markenscoff:1980,Markenscoff:1985} for a discussion on this point.

As a final remark of this subsection, we note that the earlier assumption of monotonicity of $l(t)$ can be relaxed after performing a variable transformation (from length $x'$ to elapsed time $t'$) in the remaining integral above where $x'=l(t')$ (or equivalently $t'=\eta(x')$).
This point has been discussed by Freund \cite{Freund:1973} and we leave this exercise to the reader.

\subsection{Constant velocity}
\label{sec:constvel}

The simplest case one can study within the general solution \eqref{eq:accanis} is a dislocation at rest at time $t<0$ which suddenly starts moving at constant velocity $v$ from $t\ge0$.
Then
\begin{align}
\eta(x) &= \sgn{x}\eta(\abs{x})=\frac{x}{v}
\,, & \eta'(x)&=\frac{1}{v}
\,,& \tau &= t - \left(\eta(x)-x\eta'(x)\right) = t
\,,\nn\\
\tilde{\eta} &= \frac{x'}{v}=\eta(x')
\,,\label{eq:constvel}
\end{align}
i.e. all the $\cF(\eta)$ terms cancel one another identically and the shift in time variable $\tau$ is zero allowing for further simplifications of the remaining terms:
\begin{subequations}\label{eq:steadyanis}
\begin{align}
&\partial_x u_z^{v}(x,y,t)
=\frac{by}{2\pi R^2}
\sqrt{\frac{1}{\tilde{C}} \left(1 -  \frac{\tilde{B}^2}{4\tilde{C}}  \right)}
\nn\\
&\quad +\frac{by}{2\pi R^2} \Theta\left(t-\tfrac{R}{c_A\sqrt{1-\frac{\tilde{B}^2}{4\tilde{C}}}}\right) 
 \frac{\left[vt\left(x - {y}\frac{\tilde{B}}{2\tilde{C}} \right)\frac{1}{\tilde{C}}\left[2\left(1 -  \frac{\tilde{B}^2}{4\tilde{C}}  \right)  - \frac{R^2}{\tau^2 c_A^2} 
 \right] 
 - v^2\left(t^2  - \frac{y^2}{c_A^2\tilde{C}} \right)\frac{1}{\tilde{C}}\left(1 -  \frac{\tilde{B}^2}{4\tilde{C}}  \right)\right]}{\sqrt{\frac{1}{\tilde{C}} \left(1 -  \frac{\tilde{B}^2}{4\tilde{C}}  \right) - \frac{R^2}{t^2c_A^2\tilde{C}}}\left(v^2t^2  - \frac{y^2v^2}{c_A^2\tilde{C}} - 2vt\left( x - {y}\frac{\tilde{B}}{2\tilde{C}} \right) + {R^2}\right)}
\nn\\
&\quad + \frac{b\,\sgn{y}}{2 }\sgn{{y}\frac{\tilde{B}}{2\tilde{C}} - x}\Theta\left(p_0-\frac{1}{v}\right)\delta\left(x - vt + \abs{y}\sqrt{\frac{v^2}{c_A^2\tilde{C}}-\frac{1}{\tilde{C}}\left(1-\frac{\tilde{B}^2}{4\tilde{C}}\right)}-{y}\frac{\tilde{B}}{2\tilde{C}}\right)
,\label{eq:steadyanis_a}
\end{align}
\begin{align}
&\partial_y u_z^{v}(x,y,t)
=\frac{-bx}{2\pi R^2}\sqrt{\frac{1}{\tilde{C}} \left(1 -  \frac{\tilde{B}^2}{4\tilde{C}}  \right)}
\nn\\&\quad
+\frac{b}{2\pi}
\frac{1}{R^2} \Theta\left(t-\tfrac{R}{c_A\sqrt{1-\frac{\tilde{B}^2}{4\tilde{C}}}}\right) 
 \frac{{vt}\left[\left(  \frac{y^2}{\tilde{C}}  -x^2\right)\left(1 -  \frac{\tilde{B}^2}{4\tilde{C}}  \right) + (x-vt) 
  \frac{R^2}{t^2 c_A^2}\left(x - {y}\frac{\tilde{B}}{2\tilde{C}} \right)\right] + {x}v^2\left(t^2  - \frac{y^2}{c_A^2\tilde{C}} \right)\left(1 -  \frac{\tilde{B}^2}{4\tilde{C}}  \right)}{\tilde{C}\sqrt{\frac{1}{\tilde{C}} \left(1 -  \frac{\tilde{B}^2}{4\tilde{C}}  \right) - \frac{R^2}{t^2c_A^2\tilde{C}}}\left(v^2t^2  - \frac{y^2v^2}{c_A^2\tilde{C}} - 2tv\left( x - {y}\frac{\tilde{B}}{2\tilde{C}} \right) + {R^2}\right)}
 \nn\\&\quad
 + \frac{b}{2}\sgn{{y}\frac{\tilde{B}}{2\tilde{C}} - x}\left(\sqrt{\frac{v^2}{c_A^2\tilde{C}}-\frac{1}{\tilde{C}}\left(1-\frac{\tilde{B}^2}{4\tilde{C}}\right)}-\sgn{y}\frac{\tilde{B}}{2\tilde{C}}\right)\Theta\left(p_0-\frac{1}{v}\right)
 \nn\\&\qquad\qquad\times
\delta\left(x-vt + \abs{y}\sqrt{\frac{v^2}{c_A^2\tilde{C}}-\frac{1}{\tilde{C}}\left(1-\frac{\tilde{B}^2}{4\tilde{C}}\right)}-{y}\frac{\tilde{B}}{2\tilde{C}}\right)
.\label{eq:steadyanis_b}
\end{align}
\end{subequations}
The sudden jump from static to constant motion is of course unphysical, but in the large time limit the present expression must tend to the steady state solution.
This limit must be taken carefully, taking into account that the dislocation has moved in the $x$ direction by $v\delta t$ in every time interval $\delta t$.
Hence, we introduce coordinate $x'\coleq x-vt$ which moves with the dislocation, and take $t\to\infty$ while keeping $x'$ fixed, i.e. all occurrences of $x$ must be replaced by $x=x'+vt$ prior to taking $t$ to infinity.
Using
\begin{align}
\lim_{t\to\infty}\Theta\left(t-\tfrac{R}{c_A\sqrt{1-\frac{\tilde{B}^2}{4\tilde{C}}}}\right)&=\lim_{t\to\infty}\Theta\left(t-\frac{vt}{c_A\sqrt{1-\frac{\tilde{B}^2}{4\tilde{C}}}}\right)=\Theta\left(1-\frac{v}{c_A\sqrt{1-\frac{\tilde{B}^2}{4\tilde{C}}}}\right)
\,,\nn\\
 \lim_{t\to\infty}\Theta\left(p_0-\frac{1}{v}\right) &= \Theta\left(\frac{1}{c_A\sqrt{1-\frac{\tilde{B}^2}{4\tilde{C}}}} -\frac{1}{v}\right)
 \,,
\end{align}
we find after some algebra:
\begin{subequations}\label{eq:analyticsteadystate}
\begin{align}
\lim_{t\to\infty}\partial_{x'} u_z^v(x(x',t),y,t)
&= \frac{b}{2\pi}\Theta\left(1-\frac{v}{c_A\sqrt{1-\frac{\tilde{B}^2}{4\tilde{C}}}}\right)\frac{y\sqrt{\frac{1}{\tilde{C}} \left(1 -  \frac{\tilde{B}^2}{4\tilde{C}} - \frac{v^2}{c_A^2} \right)}}{\left(x'-\frac{\tilde{B}}{2\tilde{C}}y\right)^2   + y^2\frac{1}{\tilde{C}}\left(1-  \frac{\tilde{B}^2}{4\tilde{C}}  - \frac{v^2}{c_A^2}  \right)}
\nn\\&\quad
 - \frac{b\,\sgn{y}}{2 }\delta\left(x' + \abs{y}\sqrt{\frac{v^2}{c_A^2\tilde{C}}-\frac{1}{\tilde{C}}\left(1-\frac{\tilde{B}^2}{4\tilde{C}}\right)}-{y}\frac{\tilde{B}}{2\tilde{C}}\right)
\,,\label{eq:analyticsteadystate_a}
\end{align}
\begin{align}
&\lim_{t\to\infty}\partial_y u_z^v(x(x',t),y,t)
= \frac{b}{2\pi}\Theta\left(1-\frac{v}{c_A\sqrt{1-\frac{\tilde{B}^2}{4\tilde{C}}}}\right) \frac{ - x'\sqrt{\frac{1}{\tilde{C}} \left(1 -  \frac{\tilde{B}^2}{4\tilde{C}} - \frac{v^2}{c_A^2} \right)} }{\left(x'-\frac{\tilde{B}}{2\tilde{C}}y\right)^2   + y^2\frac{1}{\tilde{C}}\left(1-  \frac{\tilde{B}^2}{4\tilde{C}}  - \frac{v^2}{c_A^2}  \right)}
\nn\\&\qquad\qquad
 -  \frac{b}{2}\left(\sqrt{\frac{v^2}{c_A^2\tilde{C}}-\frac{1}{\tilde{C}}\left(1-\frac{\tilde{B}^2}{4\tilde{C}}\right)}-\sgn{y}\frac{\tilde{B}}{2\tilde{C}}\right)
\delta\left(x' + \abs{y}\sqrt{\frac{v^2}{c_A^2\tilde{C}}-\frac{1}{\tilde{C}}\left(1-\frac{\tilde{B}^2}{4\tilde{C}}\right)}-{y}\frac{\tilde{B}}{2\tilde{C}}\right)
\,.\label{eq:analyticsteadystate_b}
\end{align}
\end{subequations}
Notice, the presence of a `critical' velocity \cite{Teutonico:1961}
\begin{align}
v_\txt{crit}&= c_A\sqrt{\left(1 -  \frac{\tilde{B}^2}{4\tilde{C}}\right)}
\,,\label{eq:vcrit}
\end{align}
which separates a `subsonic' from a `supersonic' regime.
As emphasized in Ref. \cite{Blaschke:2020MD}, $v_\txt{crit}$ is in general \emph{different} from any sound speed moving in the direction parallel to the dislocation.
In particular, the solution \eqref{eq:analyticsteadystate} diverges at the contours $x'=\frac{\tilde{B}}{2\tilde{C}}y$ when the velocity reaches $v_\txt{crit}$.

At all subsonic speeds, $v<v_\txt{crit}$
the delta functions vanish identically for all real $x'$ and $y$ since the square roots in their arguments become imaginary.
Therefore, the step function ensuring supersonic motion in those terms could be dropped above.
Integrating any of the latter two expressions with respect to $x'$ or $y$ in the strictly subsonic regime yields the solution for (subsonic) $u_z$ in the stationary limit\footnote{
A note regarding the definition of the $\arctan()$ function is in order:
We adopt here the most common definition (i.e. the IEEE convention) where $\arctan(y/x) \in [0,+\pi]$ in the first and second quadrant and $\arctan(y/x) \in [-\pi,0]$ in the third and fourth quadrant.
Ref. \cite{Hirth:1982} in contrast uses the less common definition $\arctan(y/x)=\phi\,,\quad \forall \phi\in[0,2\pi]$, and this is the reason we have an overall minus sign in front of the denominator of the argument of the arc-tangent in \eqnref{eq:uzsteadystate} rather than an overall sign in front of the $\arctan()$ like in Ref. \cite[Eq. (13-128)]{Hirth:1982}.}:
\begin{align}
\lim_{t\to\infty} u_z^v(x(x',t),y,t)
&= \frac{b}{2\pi} \arctan\left(\frac{y\sqrt{\frac{1}{\tilde{C}} \left(1 -  \frac{\tilde{B}^2}{4\tilde{C}} - \frac{v^2}{c_A^2} \right)}}{-x'+\frac{\tilde{B}}{2\tilde{C}}y}\right)
\,,&&\forall v<v_\txt{crit}
\,, \label{eq:uzsteadystate}
\end{align}
which coincides (as expected) with the solution discussed in \cite{Teutonico:1961,Blaschke:2020MD}.
Upon plugging in the expressions \eqref{eq:ABCfcc} for fcc slip systems, the static limit ($v\to0)$ of the solution above coincides with what was derived in Ref. \cite[Eq. (13-128)]{Hirth:1982}.
Note that the static limit also follows directly from \eqnref{eq:steadyanis} upon setting $v=0$, since all but the first term of both expressions are zero in this case.

\subsection[Two special cases: constant acceleration \texorpdfstring{${a}$}{a} and steadily increasing rate of \texorpdfstring{${a}$}{a}]{Two special cases: constant acceleration $\mathbf{a}$ and steadily increasing rate of $\mathbf{a}$}
\label{sec:constacc}

\begin{figure}[ht]
\centering
\includegraphics[width=0.5\textwidth]{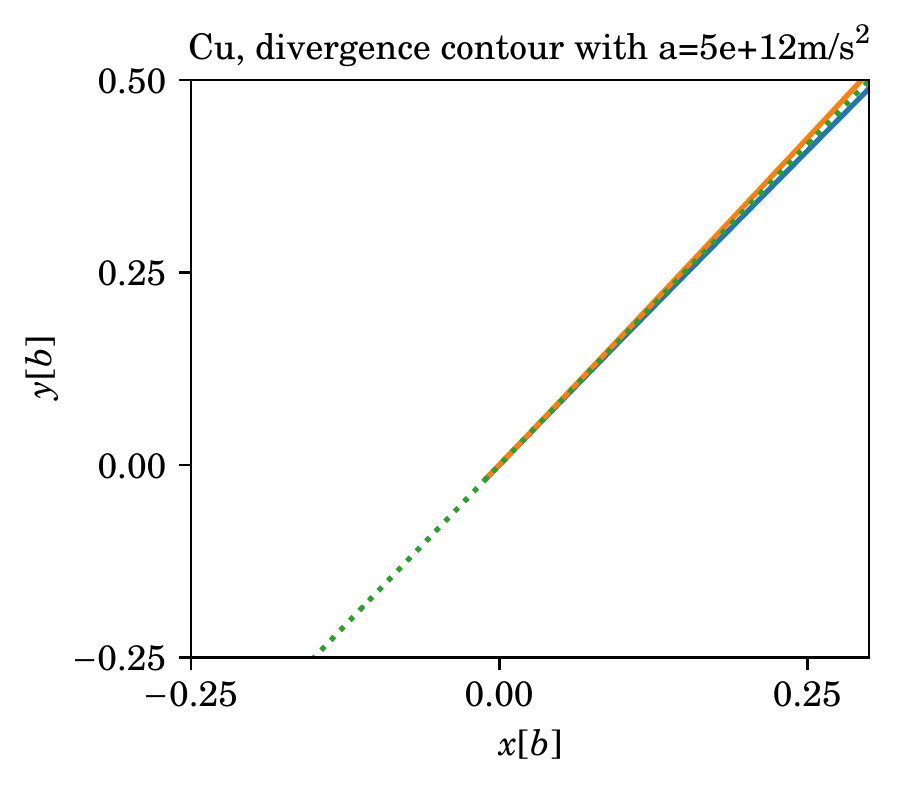}%
\includegraphics[width=0.5\textwidth]{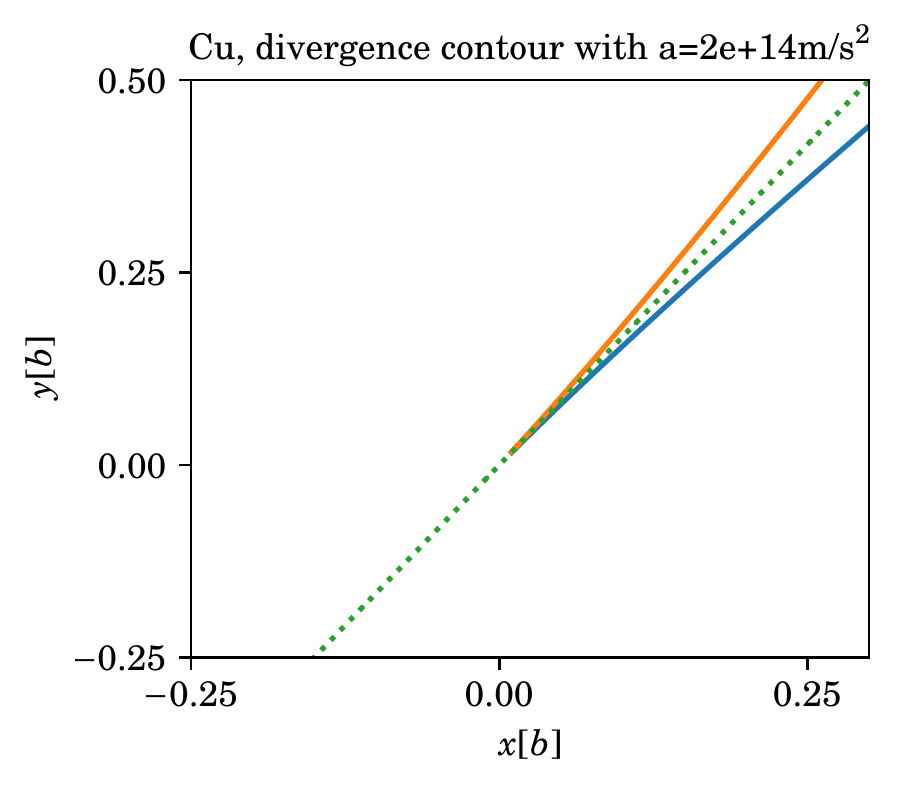}
\caption{We show the contours $y(x)$ where $\partial_xu_z$, $\partial_yu_z$ diverge when reaching the critical velocity $v(t)=v_\txt{crit}$ (solid lines) at the example of copper  ($\rho=8.96$g/ccm, $b=2.56$\r{A}, $c_{44}=75.7$GPa, and $c'=23.55$GPa, see \cite{CRCHandbook,Blaschke:2020MD}) for two different values for acceleration $a$.
For comparison, the dotted line shows what the contour looks like in the constant velocity case.
We see that for small acceleration $a$, the two solid lines almost collapse to the dotted line in the positive $x$, $y$ region, but for non-vanishing acceleration they always become complex for negative $x$.
Even though there are no real solutions for negative $x$ to equation \eqref{eq:contour_smalla}, the last term can become arbitrarily small for very small $a$ leading to a significantly enhanced dislocation field close to the dotted line (which is the solution at $a=0$) even for negative $x$.
This, of course, is expected since the general solution must tend to the constant velocity solution as $a\to0$, see also Fig. \ref{fig:comparesteadystate}.}
\label{fig:contour_smalla}
\end{figure}

\begin{figure}[ht]
\centering
 \includegraphics[width=0.5\textwidth]{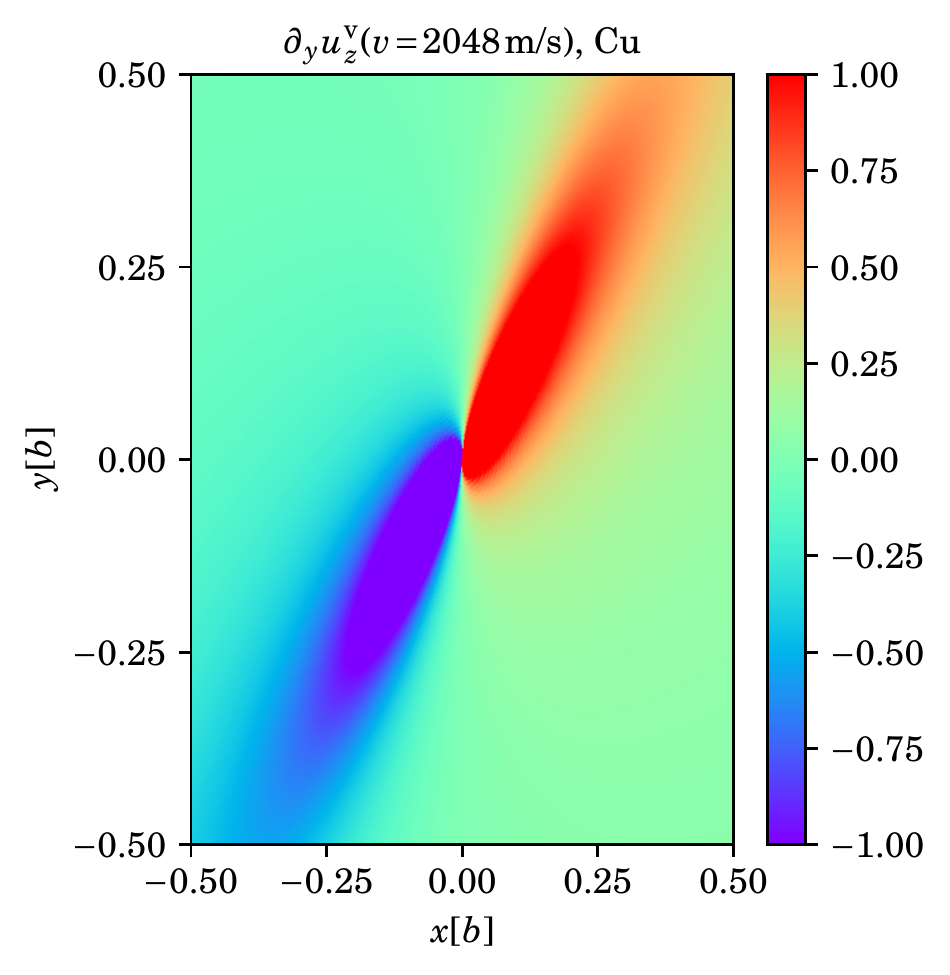}%
 \includegraphics[width=0.5\textwidth]{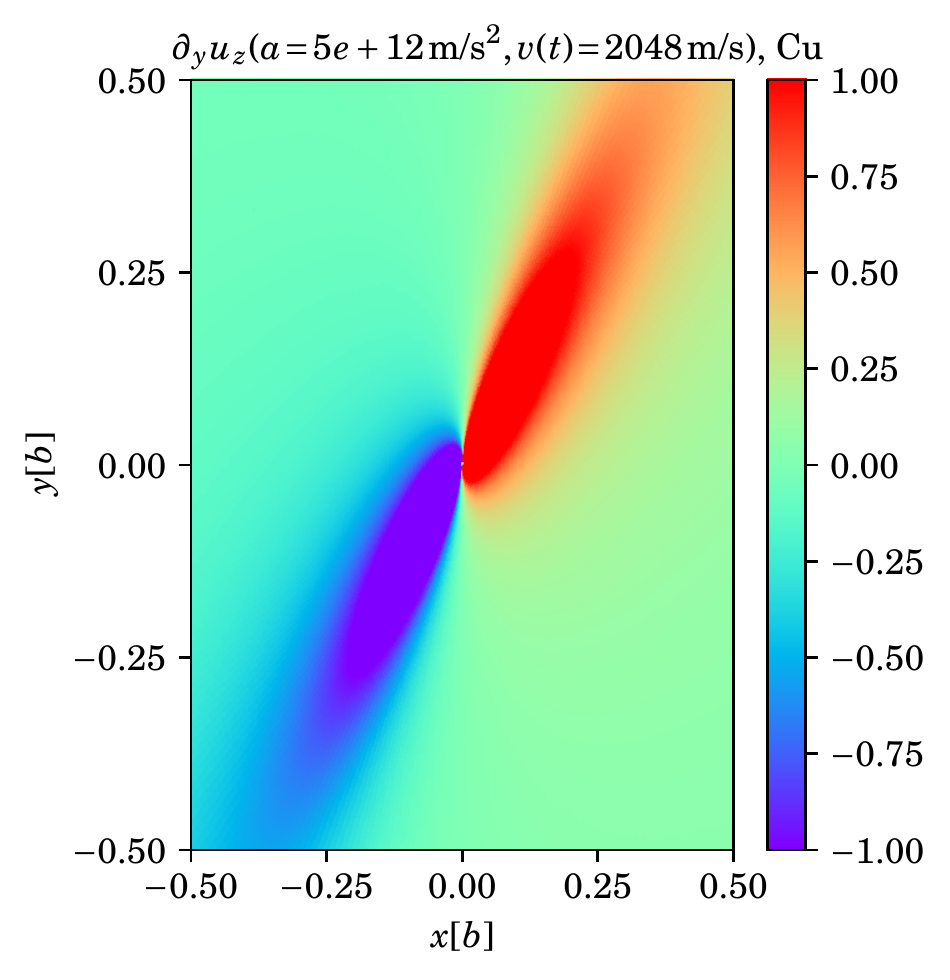}
\caption{We show $\partial_yu_{z}$ at dislocation velocity
$v=2.05\txt{km/s}$
for fcc copper ($\rho=8.96$g/ccm, $b=2.56$\r{A}, $c_{44}=75.7$GPa, and $c'=23.55$GPa, see \cite{CRCHandbook,Blaschke:2020MD}).
This velocity corresponds to roughly 93\% of the critical velocity and coincides with the lowest shear wave speed propagating in the direction of $v$,
and as argued in Ref. \cite{Blaschke:2020MD}, nothing special happens at this velocity.
Both plots are centered at the dislocation core, showing the plane perpendicular to the dislocation line in units of a Burgers vector.
On the left, we show the steady state-solution \eqref{eq:analyticsteadystate_b} and on the right we show the full solution for constant acceleration \eqref{eq:accanis_b} with \eqref{eq:constacc} and $a=5\times10^{12}$m/s$^2$ at time $t_v= v/a = 4.1\times10^{-10}$s needed to reach velocity $v$.
At this point, the dislocation has traveled a distance of $0.42$ microns.
The integration w.r.t. $x'$ in the full solution as well as the subsequent time derivative have been carried out numerically, though we point out that the $\cF$ dependent terms are much smaller compared to the others and the figure on the right does not change visibly if those terms are neglected.
We see that the changes in the dislocation displacement gradient due to the inclusion of acceleration are barely visible in this example.}
\label{fig:comparesteadystate}
\end{figure}

\begin{figure}[ht]
\centering
\includegraphics[width=0.51\textwidth]{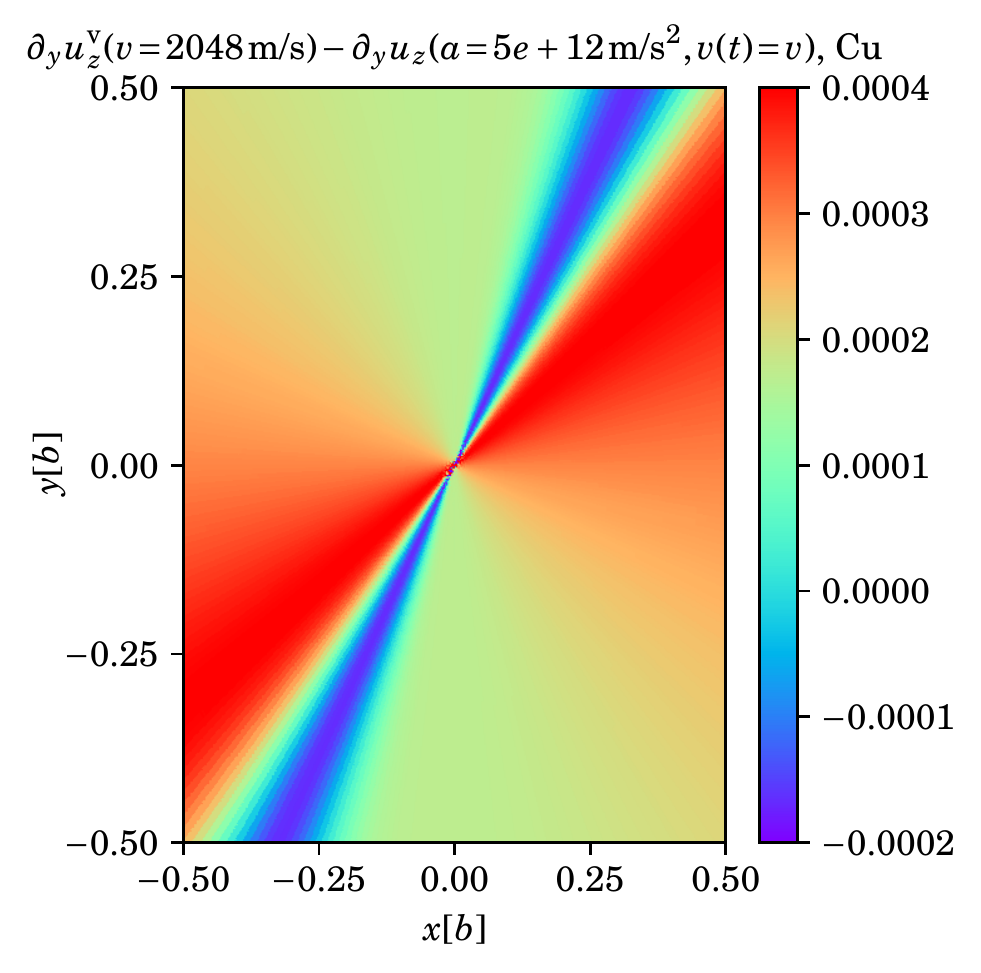}%
\includegraphics[width=0.49\textwidth]{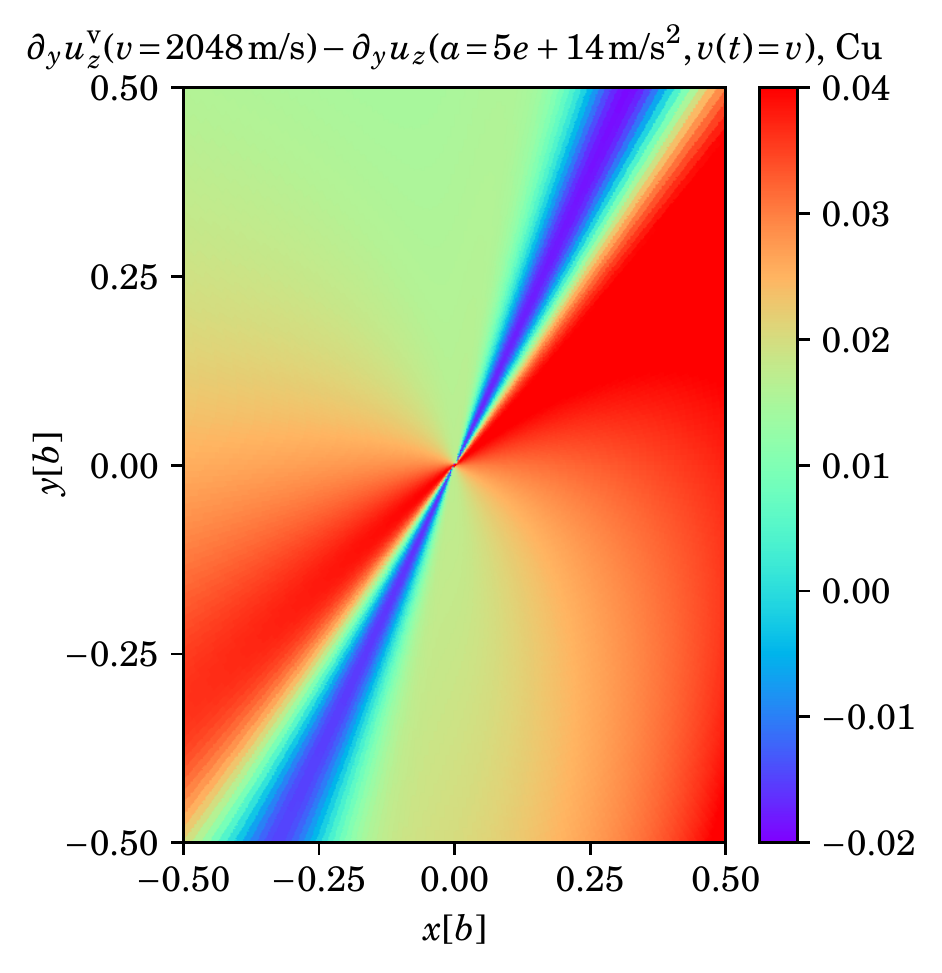}
\caption{We show the difference of the steady state solution and the dynamic solution at time $t$ when $v(t)$ matches the velocity of the steady-state solution, i.e. $\partial_yu^\text{v}_{z}-\partial_yu_{z}(a,v(t)=v)$, once more for fcc copper using the same parameters as in Figure \ref{fig:comparesteadystate}.
On the left, we show this difference for an acceleration of $a=5\times10^{12}$m/s$^2$.
As such, this figure corresponds to the difference of the two Sub-figures  \ref{fig:comparesteadystate}.
On the right, we repeat this exercise for an (extreme)  hundred-fold faster acceleration of $a=5\times10^{14}$m/s$^2$.
Note the different ranges in the color bar:
The average magnitude of this difference scales almost proportional to the acceleration, i.e. the color bar range increased a hundredfold for a hundredfold increase in $a$, though some smaller differences in the patterns are also clearly visible (especially in the bottom right corner of this example).
}
\label{fig:comparesteadystateacc}
\end{figure}

\begin{figure}[ht]
\centering
\includegraphics[width=0.51\textwidth]{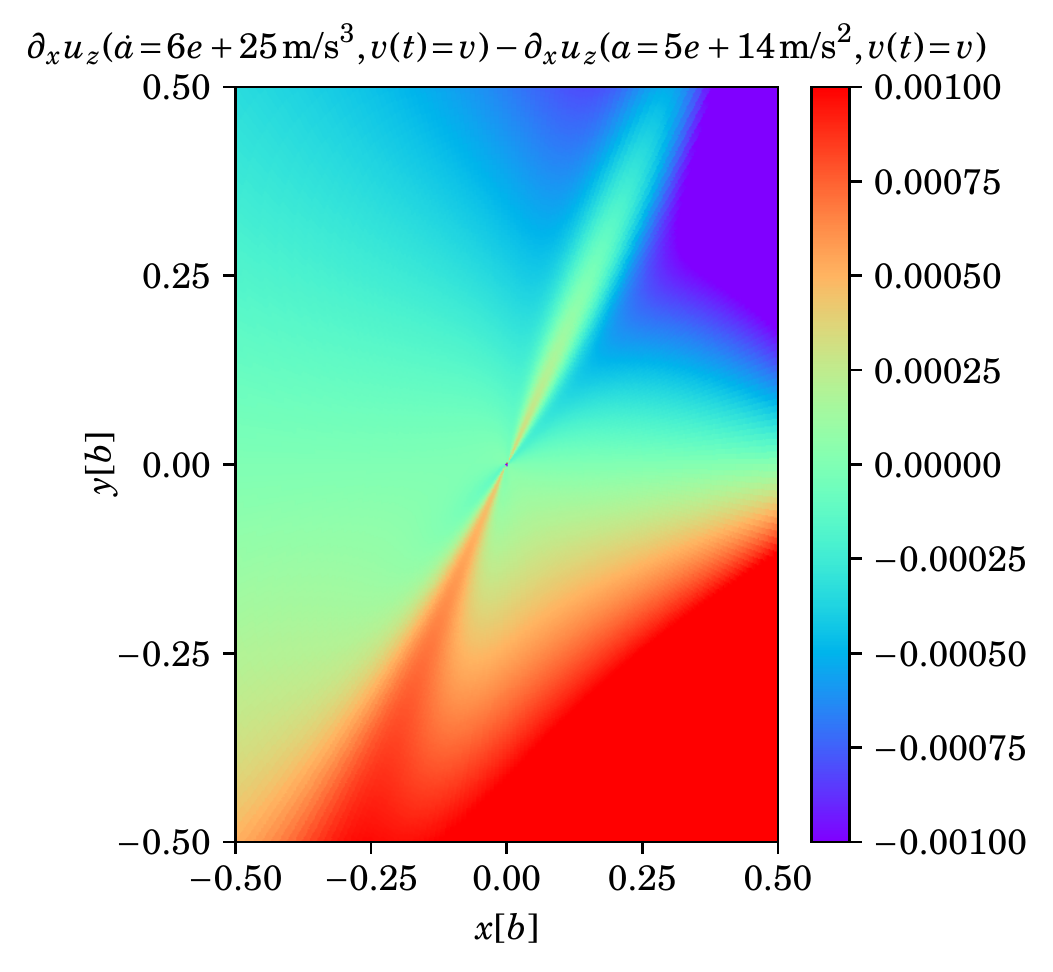}%
\includegraphics[width=0.49\textwidth]{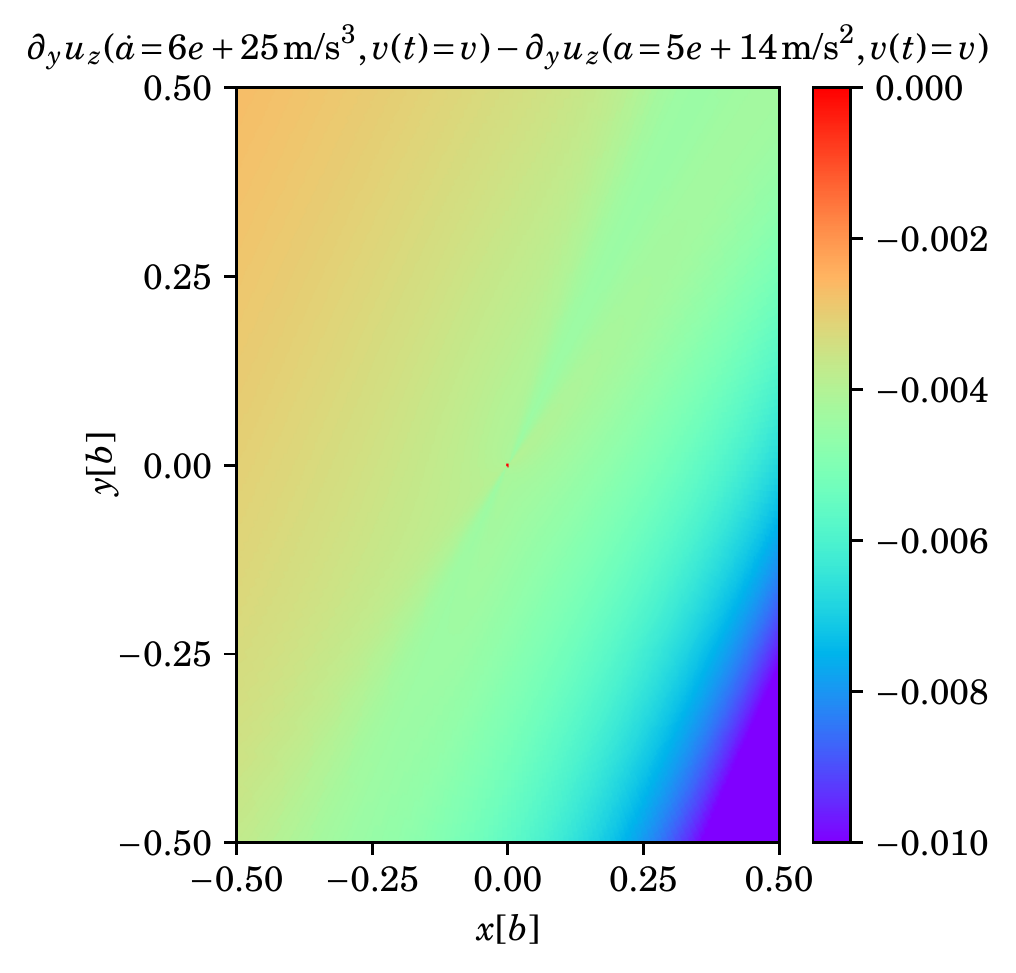}
\caption{We show the difference between the dynamic solution with constant acceleration $a=5\times10^{14}$m/s$^2$ and the dynamic solution assuming $a(t=0)=0=v(t=0)$ and $\partial_t a(t>0)=6.2\times10^{25}$m/s$^3$, both for fcc Cu at times $t$ when their respective velocities $v(t)$ match the velocity $v=2048$m/s.
The increase in acceleration $\dot a=6.2\times10^{25}$m/s$^3$ was chosen such that $a(t)=5\times10^{14}$m/s$^2$ and $v(t)=2048$m/s at target time $t$.
All other parameters are the same as in Figures \ref{fig:comparesteadystate} and \ref{fig:comparesteadystateacc}.
We see that including a change in acceleration with time leads to an even smaller correction to the solution than the inclusion of acceleration had compared to the steady-state solution.
}
\label{fig:compareacc}
\end{figure}

\begin{figure}[ht]
\centering
\includegraphics[width=0.475\textwidth]{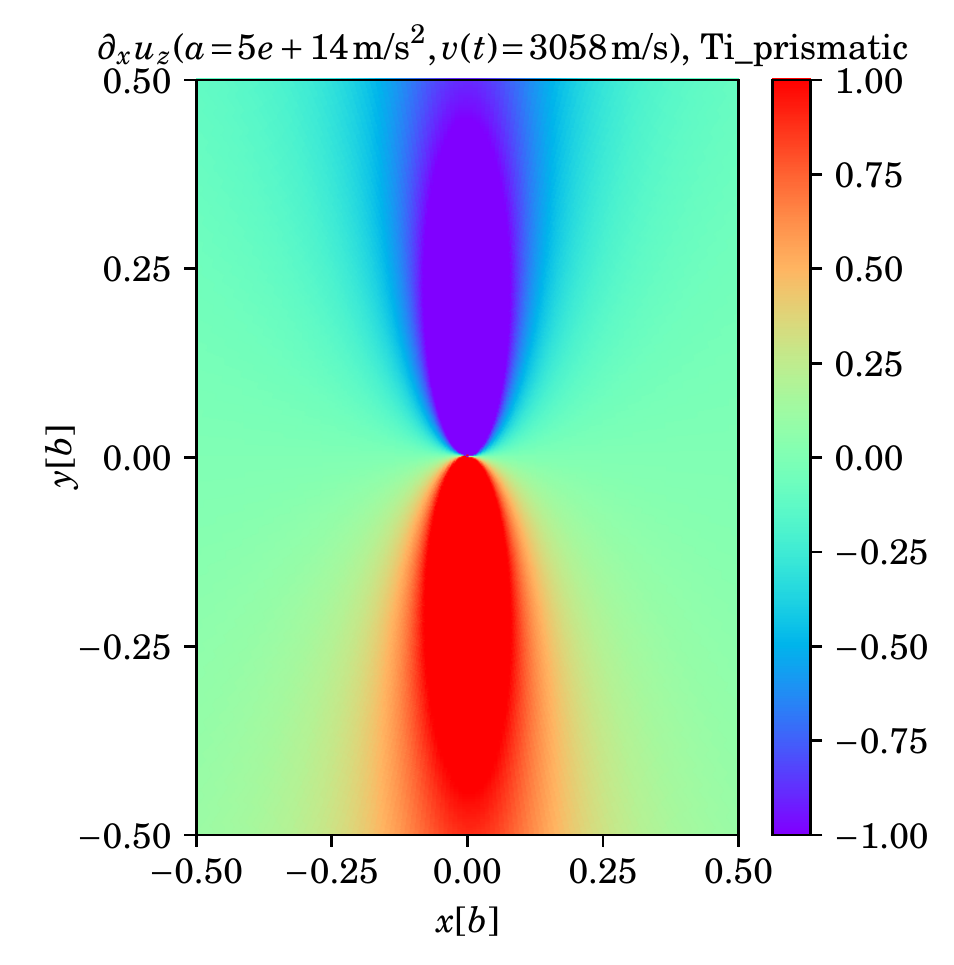}%
\includegraphics[width=0.525\textwidth]{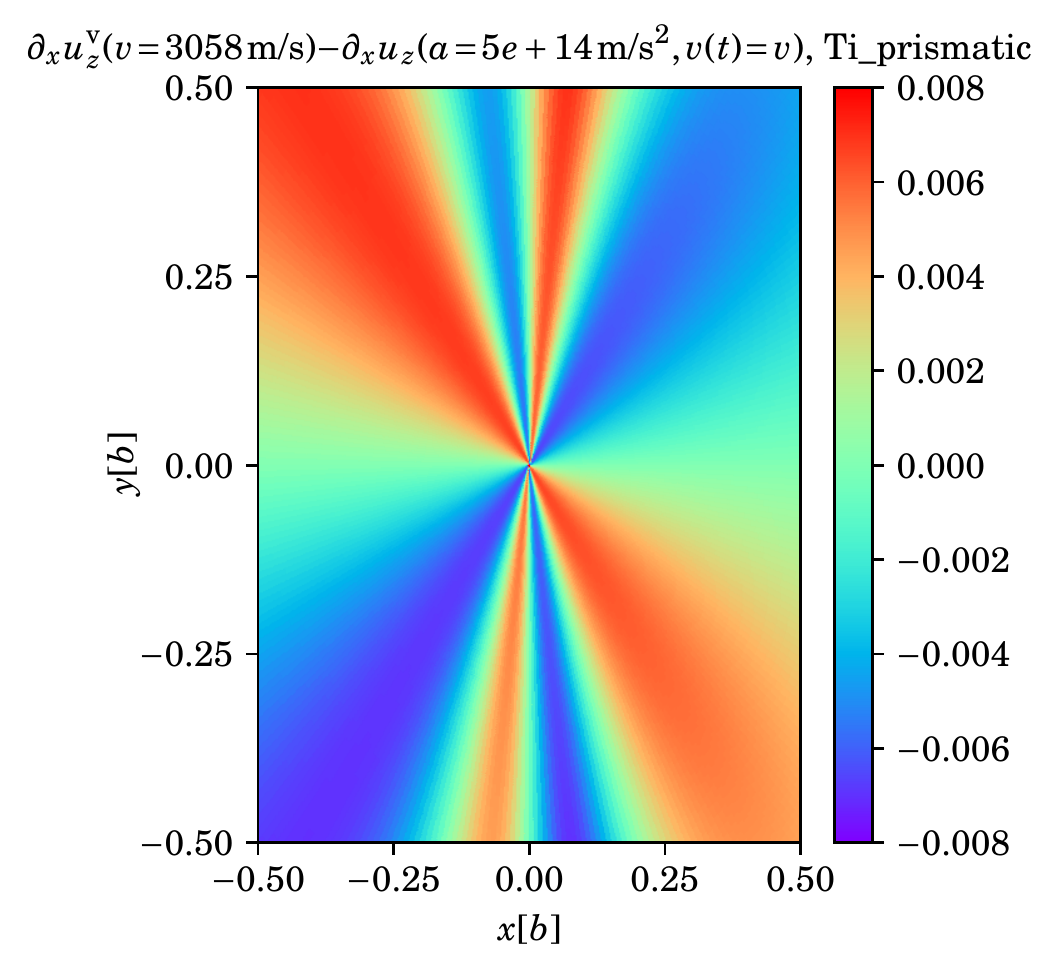}
\caption{We show as another example a component of the displacement gradient of an accelerating screw dislocation in hcp titanium ($\rho=4.506$g/ccm, $a=2.9506$\r{A}, $c=4.6835$\r{A}, $c_{11}=162.4$GPa, $c_{12}=92.0$GPa, $c_{13}=69.0$GPa, $c_{33}=180.7$GPa, and $c_{44}=46.7$GPa, see \cite{CRCHandbook,Blaschke:2019fits}).
Its Burgers vector is  [$\bar{2},1,1,0$] and its glide plane is ($\bar{1},0,1,0$).
On the left, we show one component of strain at time $t$ when the dislocation reaches 95\% of its critical velocity after being accelerated at a constant rate of $a=5\times10^{14}$m/s$^2$.
On the right, we show the difference between this solution and the steady state case at the same dislocation velocity.
}
\label{fig:titanium}
\end{figure}

Let us now assume again the dislocation is at rest at time $t<0$ and starts to accelerate at a constant rate $a$ from $t\ge0$.
Then  $l(t)=\frac{a}{2}t^2>0$ and hence
\begin{align}
\eta(x) &= \sgn{x}\sqrt{\frac{2\abs{x}}{a}}
\,, & \eta'(x)&
=\sgn{x}\partial_x\eta(\abs{x})
=\frac{\eta(x)}{2x}
\,,& \tau &= t - \left(\eta(x)-x\eta'(x)\right)
 = t - \frac{1}{2}\eta(x)
\,,\nn\\
\tilde{\eta} &= 
\frac{1}{2}\left(1+\frac{x'}{x}\right)\eta(x)
\,. \label{eq:constacc}
\end{align}
Because the current velocity is given by $v(t)=at$,
the transition from subsonic to supersonic in the above solution happens when $t=v_\txt{crit}/a$.
At this point, the dislocation has traveled a distance $x=v_\txt{crit}^2/(2a)$, and it is at this position where it is most obvious that the core-singularity at $y=0$ is enhanced within $\partial_yu_z$.
Furthermore, like in the constant velocity case there are contours where the solution diverges when the critical velocity is reached.
Those contours follow from\footnote{We see immediately, that in the constant velocity limit the second term is zero as $\eta'=1/v\to 1/v_\txt{crit}$, and the first term vanishes for the contour $(x-vt)=y\frac{\tilde{B}}{2\tilde{C}}$ since $\tau=t$ in that case.}
\begin{align}
0&=\tau^2-\frac{y^2}{c_\txt{A}^2\tilde{C}}-2\tau\eta'\left(x-y\frac{\tilde{B}}{2\tilde{C}}\right)+\eta'^2R^2
\nn\\
&=\left[\tau-\eta'\left(x-y\frac{\tilde{B}}{2\tilde{C}}\right)\right]^2+\left(\eta'^2-\frac{1}{v_\txt{crit}^2}\right)\frac{y^2}{\tilde{C}}\left(1-\frac{\tilde{B}}{4\tilde{C}}\right)
\,,
\end{align}
see the denominators in \eqnref{eq:accanis}.
Real solutions $y(x)$ can only be expected for $\eta'v_\txt{crit}\le1$, and in proximity to the dislocation core this is the case when the dislocation velocity $v(t)$ approaches $v_\txt{crit}$.
It will be convenient to shift $x=\to x+at^2/2$ so that $x=0=y$ is the current position of the dislocation core.
Consider now $\tau=t-(x+at^2/2)\eta'$,  $t=v_\txt{crit}/a$, and $\eta'^2=1/\abs{2ax+v_\txt{crit}^2}$.
Close to the core, where the product $2a\abs{x}<v_\txt{crit}^2$, we may drop the absolute value in $\eta'$.
Hence, dividing the whole equation above by $\eta'^2$ we presently have 
\begin{align}
0
&=\left[\frac{v_\txt{crit}}{a\eta'}-\left(2x+\frac{v_\txt{crit}^2}{a}\right) +y\frac{\tilde{B}}{2\tilde{C}}\right]^2+\left(1-\frac{1}{v_\txt{crit}^2\eta'^2}\right)\frac{y^2}{\tilde{C}}\left(1-\frac{\tilde{B}}{4\tilde{C}}\right)
\nn\\
&=\left[\frac{v_\txt{crit}}{a}\sqrt{2ax+v_\txt{crit}^2}-\frac{1}{a}\left(2ax+{v_\txt{crit}^2}\right) +y\frac{\tilde{B}}{2\tilde{C}}\right]^2 - \frac{2ax}{v_\txt{crit}^2}\frac{y^2}{\tilde{C}}\left(1-\frac{\tilde{B}}{4\tilde{C}}\right)
\,, \label{eq:contour_smalla}
\end{align}
which is a quadratic equation for $y(x)$ with real solutions only for positive $x$.
Typically, critical velocities are of the order of a few km/s and Burgers vector length scales are a few \r{A}ngstr\"om.
Therefore, if $v_\txt{crit}\sim 10^3$m/s and $x<10^{-9}$m near the core, $2ax<v_\txt{crit}^2$ implies $a<10^{15}$m/s$^2$ as an upper limit for dropping the absolute value in $\eta'^2$ leading to \eqref{eq:contour_smalla}.
Figure \ref{fig:contour_smalla} illustrates the solutions to this equation for two different values of $a$ as well as for $a=0$ (which is the only case where a real solution exists for $x<0$).
In the isotropic limit, $\tilde{B}\to0$ (see Section \ref{sec:isotropic} below), and the contours become $y(x)\approx \frac{x v_\txt{crit}^2}{2a}+\cO(x^2)\ll1$ for positive $x$ and $a\ne0$, and $x=0$ for any $a$ (including zero).
%

Markenscoff et al. \cite{Markenscoff:2008} argue that independent of $\eta(x)$, the singularities in the dislocation field are only removed once the core itself is regularized (i.e. modeled to be of finite size) in some fashion; see also \cite{Callias:1980,Callias:1988} for a discussion of the wave-front asymptotics in the isotropic limit.
Similarly, the author of Ref. \cite{Pellegrini:2018} shows that a regularized core can remove the singularity in the anisotropic steady-state special case, see \eqref{eq:analyticsteadystate}.

Another point to note is that for small accelerations $a$, the $\cF$ dependent terms are small compared to the others (and vanish identically for constant velocity as seen in the previous subsection), and these terms become important only when dislocations accelerate quickly.
In dislocation dynamics where, according to MD simulations, dislocations reach their steady state velocity within picoseconds (see e.g. \cite{Blaschke:2020MD}), ``small'' accelerations can still mean of the order of $a\sim10^{12}$m/s$^2$.

Figure \ref{fig:comparesteadystate} shows an example of a screw dislocation in copper moving at about 93\% $v_\txt{crit}$, once for the steady state solution \eqref{eq:analyticsteadystate}, and once for the full solution \eqref{eq:accanis_b} with constant acceleration $a=5\times10^{12}$m/s$^2$, see \eqref{eq:constacc}
(as well as the open source code PyDislocDyn \cite{pydislocdyn} for its numerical implementation).
The acceleration $a$ was chosen at a typical peak value according to single crystal plasticity simulations \cite{Blaschke:2021impact} and high enough such that the dislocation can easily achieve its target velocity by traveling less than a micron, i.e. in a fraction of even the smallest single crystal grains within a polycrystal.
Furthermore, ``low'' dislocation densities within a single crystal are considered to be of the order of $10^6$mm$^{-2}$ which means the mean free path of an accelerating dislocation will then also be of the order of one micron \cite{Austin:2018,Blaschke:2019a}.

We see that, indeed as expected, the dislocation field of a moderately accelerated screw dislocation at any given time-snapshot does not look too different from the steady state solution \eqref{eq:analyticsteadystate} at the same velocity.
Figure \ref{fig:comparesteadystateacc} quantifies this statement by showing that the differences in strain are less than 0.001 for an acceleration of $a=5\times10^{12}$m/s$^2$.
In the somewhat extreme case of e.g. a hundred fold larger acceleration of $a=5\times10^{14}$m/s$^2$, the differences in strain also increase roughly a hundred fold.
Moving even closer to the dislocations critical velocity also enhances the effect of acceleration.
In other words, the dynamic solution becomes increasingly important for accelerations $a\gtrsim10^{14}$m/s$^2$ for fast moving dislocations.
As such, the steady state solution \eqref{eq:analyticsteadystate} can be expected to be a good (and computationally far less expensive) approximation for subsonic screw dislocations within, say discrete dislocation dynamics (DDD) or single crystal plasticity simulations, except for extreme situations of dislocations being accelerated very fast up to speeds close to their critical velocity.
To solidify this statement, Figure \ref{fig:compareacc} shows numerical results for the dynamic test case of a screw dislocation being accelerated at a steadily \emph{increasing} rate $a=\dot a t$, i.e.
\begin{align}
	l(t) &= \dot a\frac{t^3}{6}\,, \qquad
	\eta(x) = \sgn{x}\sqrt[3]{\frac{6\abs{x}}{\dot a}}\,,\qquad
	\eta'(x) = \frac{\eta(x)}{3x}
	\,,\nn\\
	\tilde\eta&=\eta(x)+(x'-x)\eta'(x) =\frac{1}{3}\left(2+\frac{ x'}{x}\right)\eta(x)
	\,,\qquad \tau = t - \left(\eta(x)-x\eta'(x)\right) =  t - \frac23 \eta(x)
	\,,
\end{align}
where $\dot a\equiv\partial_t a(t)$ is assumed constant and positive from time $t>0$ and $a(t=0)=0=v(t=0)$.
We see that including a change in acceleration yields an even smaller additional correction to the steady-state solution.

Edge and mixed dislocations have yet to be studied within the present framework before we can generalize this statement.
In a real crystal we cannot expect $a$ (or even $\dot a$) to be constant, but rather $a(t)$ will initially be large and then tend to zero as the dislocation approaches its steady state velocity (if it has not encountered an obstacle before that time).

In Figure \ref{fig:titanium} we show additional results from screw dislocations on a prismatic slip plane in hcp titanium, thus emphasizing that our solution is applicable to a number of slip systems beyond those of fcc crystals.
In particular, we considered a screw dislocation with Burgers vector  [$\bar{2},1,1,0$] gliding on the slip plane ($\bar{1},0,1,0$) which can be checked to fulfill the symmetry requirements of our present solution.
Once more, the difference between the dynamic and the steady state solution is moderately small (shown this time for $\partial_xu_z$) except for extreme accelerations and dislocation velocities close to the critical velocity.

As a final remark of this subsection, we note that due to the divergence at $v(t)=v_\txt{crit}$ within \eqref{eq:accanis}, the transition to supersonic speeds will depend on how the dislocation core is modeled,
since the remaining divergence at the critical velocity can only be expected to be removed by an extended core \cite{Markenscoff:2008,Pellegrini:2018}.
While Ref. \cite{Markenscoff:2008} studied a ramp-like core to regularize the divergence only in the isotropic limit and Ref. \cite{Pellegrini:2018} studied an elliptic core in the anisotropic steady-state limit, there is no immediate reason why these strategies should not work also in the present general anisotropic case.
This, however, is beyond the scope of the present work and is left for future studies.

\subsection{The isotropic limit}
\label{sec:isotropic}

In the isotropic limit, $c_{11}=c_{12}+2c_{44}$ which leads to $\tilde{B}=0$, $\tilde{C}=C/A=1$, and $c_A=\ct=\sqrt{c_{44}/\rho}$ is the transverse sound speed.
Furthermore, $p_0=\frac{x}{r\ct}$.
The general solution \eqref{eq:accanis} then simplifies to
\begin{subequations}\label{eq:geniso}
\begin{align}
\partial_x u_z^\txt{iso}(x,y,t)
&=\frac{by}{2\pi r^2}
+\frac{b}{2\pi}\partial_t\int\limits_{0}^\infty dx' 
\left(\cF_x\left[\eta(x'),t\right] - \cF_x\left[\eta(x) + (x'-x)\eta'(x),t\right]\right)
\nn\\
&\quad +\frac{b}{2\pi}
\frac{y}{r^2}
\left[ \frac{ \Theta\left(\tau-\tfrac{r}{\ct}\right) \left(\tau x\, \eta'(x)\left[2  - \frac{r^2}{\tau^2 \ct^2} 
 \right] 
 - \left[\tau^2  - \frac{y^2}{\ct^2} \right]\right)}{\sqrt{1 - \frac{r^2}{\tau^2\ct^2}}\left(\tau^2  - \frac{y^2}{\ct^2} - {2\tau}x\,\eta'(x) + (\eta'(x))^2{r^2}\right)}\right]
\nn\\
&\quad + \frac{b\,\sgn{y}}{2 }\sgn{- x}\eta'(x)\Theta\left(\frac{x}{r\ct}-\eta'(x)\right)\,\delta\left(t - {x}\eta'(x) - \abs{y}\sqrt{\frac{1}{\ct^2}-(\eta'(x))^2}\right)
,\label{eq:geniso_a}
\end{align}
\begin{align}
\partial_y u_z^\txt{iso}(x,y,t)
&=\frac{-bx}{2\pi r^2}
+\frac{b}{2\pi}\partial_t\int\limits_{0}^\infty dx' 
\left(\cF_y\left[\eta(x'),t\right] - \cF_y\left[\eta(x) + (x'-x)\eta'(x),t\right]\right)
\nn\\&\quad
+\frac{b}{2\pi}
\frac{1}{r^2}  \Theta\left(\tau-\frac{r}{\ct}\right) 
 \frac{{\tau \eta'(x)}\left(  {y^2}  -x^2\right) + x\left(\frac{x\eta'(x)}{\tau}-1\right) 
  \frac{r^2}{\ct^2} + {x}\left(\tau^2  - \frac{y^2}{\ct^2} \right)}{\sqrt{ 1 - \frac{r^2}{\tau^2\ct^2}}\left(\tau^2  - \frac{y^2}{\ct^2} - {2\tau}x\,\eta'(x) + (\eta'(x))^2{r^2}\right)}
 \nn\\&\quad
 + \frac{b}{2}\sgn{ - x}\Theta\left(\frac{x}{r\ct}-\eta'(x)\right)\sqrt{\frac{1}{\ct^2}-(\eta'(x))^2}
\delta\left(t - {x}\eta'(x) - \abs{y}\sqrt{\frac{1}{\ct^2}-(\eta'(x))^2}\right)
,\label{eq:geniso_b}
\end{align}
\end{subequations}
with
\begin{subequations}
\begin{align}
\cF_x\left[\eta,t\right]
&=\Theta\left(t-\eta-\frac{r'}{\ct}\right)
\frac{ y\left(x-x' \right)\left(2\left(t-\eta\right)^2-\frac{r'}{\ct}\right)}{r'^4\sqrt{\left(t-\eta\right)^2 - \frac{r'^2}{\ct^2}}}
,
\end{align}
\begin{align}
\cF_y\left[\eta,t\right]
&=\Theta\left(t-\eta-\frac{r'}{\ct}\right)
\frac{1}{r'^4}\left[\frac{y^2\left(t-\eta\right)^2}{\sqrt{\left(t-\eta\right)^2 - \frac{r'^2}{\ct^2}}}
 - (x-x')^2\sqrt{\left(t-\eta\right)^2 - \tfrac{r'^2}{\ct^2}}\right]
\,,
\end{align}
\end{subequations}
and
\begin{align}
r^2 &= \left(x^2  + {y^2} \right)
\,,&
r'^2 &= \left((x-x')^2   + {y^2} \right)
\,,&
\tau &= t - \left(\eta(x)-x\eta'(x)\right) 
\,.
\end{align}
The special case of constant velocity, using relations \eqref{eq:constvel} yields
\begin{subequations}
\begin{align}
\partial_x u_z^{v,\txt{iso}}(x,y,t)
&=\frac{by}{2\pi r^2}\left[1
 +
 \Theta\left(t-\frac{r}{\ct}\right)\frac{ vt \left[x \left(2  - \frac{r^2}{t^2 \ct^2} 
 \right) 
 - vt\left(1  - \frac{y^2}{t^2\ct^2} \right)\right]}{\sqrt{1 - \frac{r^2}{t^2\ct^2}}\left((x-vt)^2  +y^2\left(1 - \frac{v^2}{\ct^2}\right)\right)}\right]
\nn\\
&\quad + \frac{b\,\sgn{y}}{2 }\sgn{ - x}\Theta\left(\frac{x}{r\ct}-\frac{1}{v}\right)\delta\left(x - vt + \abs{y}\sqrt{\frac{v^2}{\ct^2}-1}\right)
,\label{eq:steadyiso_a}
\end{align}
\begin{align}
\partial_y u_z^{v,\txt{iso}}(x,y,t)
&=\frac{b}{2\pi r^2}\left[-x + \Theta\left(t-\frac{r}{\ct}\right) 
 \frac{{vt}\left[  {y^2}  -x^2 + x\left({x}-vt\right) 
  \frac{r^2}{t^2\ct^2} + {x}vt\left(1  - \frac{y^2}{t^2\ct^2} \right)\right]}{\sqrt{ 1 - \frac{r^2}{t^2\ct^2}}\left((x-vt)^2  +y^2\left(1 - \frac{v^2}{\ct^2}\right)\right)}\right]
 \nn\\&\quad + \frac{b}{2}\sgn{ - x}\left(\sqrt{\frac{v^2}{\ct^2}-1}\right)\Theta\left(\frac{x}{r\ct}-\frac{1}{v}\right)
\delta\left(x-vt + \abs{y}\sqrt{\frac{v^2}{\ct^2}-1}\right)
,\label{eq:steadyiso_b}
\end{align}
\end{subequations}
which generalizes the earlier result of Ref. \cite[Eq. (16)]{Markenscoff:1980} for $\partial_y u_z^{v}(x,y,t)$ to all velocities including the supersonic regime.
As before, the steady-state solution follows by taking the large time limit while keeping $x'\coleq x-vt$ at a fixed value:
\begin{align}
\lim_{t\to\infty}\partial_{x'} u_z^{v,\txt{iso}}(x(x',t),y,t)
&= \frac{b}{2\pi}\frac{y\,\Theta\left(1-\abs{\beta}\right)}{(x')^2\gt   + y^2/\gt}
- \frac{b\,\sgn{y}}{2 }\delta\left(x' + \abs{y}\sqrt{\beta^2-1}\right)
\,,\nn\\
\lim_{t\to\infty}\partial_y u_z^{v,\txt{iso}}(x(x',t),y,t)
&= \frac{b}{2\pi} \frac{ - x' \,\Theta\left(1-\abs{\beta}\right)}{(x')^2\gt   + y^2/\gt}
- \frac{b}{2}\left(\sqrt{\beta^2-1}\right)
\delta\left(x' + \abs{y}\sqrt{\beta^2-1}\right)
\,,\label{eq:isotropicsteadystate}
\end{align}
where $\gt^2= 1/(1-\bt^2)$ and $\bt=v/\ct$, and the argument of the step function in the first term tends to $\pm\infty$ depending on whether $v$ is smaller or larger than the transverse sound speed.
In the subsonic regime, $v<\ct$, the delta functions are identically zero and these expressions indeed coincide with the well-known steady-state solution of Eshelby \cite{Eshelby:1949,Weertman:1980} for screw dislocations moving subsonically.
On the other hand, in the supersonic regime, $v>\ct$, the step function is zero leaving only the delta functions as the supersonic solution, consistent with the discussion in Ref. \cite[Sec. 4.2]{Weertman:1980}.

\section{Conclusion}
\label{sec:conclusion}

In this paper, we have derived the most general solution for an accelerating pure screw dislocation in an anisotropic crystal.
The term `pure' screw entails an important restriction:
Unless the slip system exhibits a `reflection' symmetry, screw and edge dislocations mix and cannot be separated.
Hence, the present solution applies only to reflection symmetric slip systems, such as all 12 fcc slip systems and a number of hcp, orthorhombic, and other slip systems, but none of the 48 bcc slip systems.

We confirm for the anisotropic screw dislocation what Markenscoff et al. have concluded some time ago for its isotropic limit:
that the divergence at a `critical' dislocation velocity (which separates the subsonic from the supersonic regime), persists for general accelerating dislocations with vanishing core size \cite{Markenscoff:2008}.
In other words, the remaining singularities (including at the core) must be removed by regularizing the core in an appropriate fashion \cite{Markenscoff:2001}.
Recent work on modeling dislocation cores (from theory) in a realistic way can be found in \cite{Clouet:2011a,Szajewski:2017phil,Pellegrini:2018,Boleininger:2019,Gurrutxaga:2019} and references therein.
Furthermore, we saw that when simulating dislocations in larger codes, the computationally less expensive steady state solution \eqref{eq:analyticsteadystate} can be expected to be a fairly good approximation for subsonic screw dislocations with low to moderate acceleration compared to its more general counterpart (and our main result) \eqnref{eq:accanis}.
The general solution \eqnref{eq:accanis} has been derived for the first time within this work,
and will also be an important starting point when studying potential transitions to supersonic speeds within future work.
The latter will depend on how the dislocation core is modeled (which is beyond the scope of this work), and will thus generalize to accelerating dislocations what has been done in Ref. \cite{Pellegrini:2018} for the steady-state limit.

Finally, we showed how various limits reduce to known results:
the isotropic limit as well as steady-state solutions for both subsonic and supersonic regimes are recovered from our general result which itself applies to any time-dependent velocity $v(t)\neq v_\txt{crit}$, as long as it does not exactly coincide with the critical velocity (which in turn is higher than the lowest shear wave speed for fcc screw dislocations \cite{Blaschke:2020MD}).

\subsection*{Acknowledgements}
\noindent
The author would like to thank B. A. Szajewski, S. Fensin, D. J. Luscher, R. G. Hoagland, and J. Chen for related discussions.
I also thank X. Markenscoff for some clarifying comments on the Laplace transform method and for drawing my attention to Refs. \cite{Markenscoff:1984,Markenscoff:1985a}.
Finally, I am equally grateful to the anonymous referees for their very insightful comments.

This work was mostly supported by the Institute for Material Science at Los Alamos National Laboratory.
In particular, the author acknowledges support by the IMS Rapid Response program.
Furthermore, the author is grateful for the support of the Materials project within the Advanced Simulation and Computing, Physics and Engineering Models Program of the U.S. Department of Energy under contract 89233218CNA000001 in the final stages of this work.

\appendix
\section{Rotation matrix and coefficients of the differential equation}
\label{sec:rotmat}

In this short appendix we review how to derive the coefficients $A$, $B$, and $C$ of the differential equation \eqref{eq:diffeq_screw_gen} for a given slip system at the example of fcc metals with Burgers unit vector $\hat{b}=(1,1,0)/\sqrt{2}$ and slip plane normal $\hat{n}_0=(1,-1,1)/\sqrt{3}$ in Cartesian coordinates.
For a screw dislocation, the line sense is parallel (or antiparallel) to $\hat{b}$, so from $\hat{t}(\vth) = \frac1b\left[\vec{b}\cos\vth+\vec{b}\times\hat{n}_0\sin\vth\right]$ with character angle $\vth=0$ we presently have $\hat{t}(0) = \hat{b}=\frac1b\vec{b}$.
Assuming a straight dislocation that is much longer than its Burgers vector length, the only velocity component that matters is the one perpendicular to the dislocation line, i.e. $\vec{v} = \pm v \hat{v}$ and $\hat{v} = \hat{n}_0\times\hat{t}(0)= \hat{n}_0\times\hat{b} = (-1,1,2)/\sqrt{6}$.
In order to derive $A$, $B$, and $C$ of the differential equation, we need the rotation matrix that aligns $\hat{b}\|\hat{z}$, $\hat{n}_0\|\hat{y}$, and $\hat{v}\|\hat{x}$.
Given a rotation axis unit vector $\hat{a}$ and an angle $\phi$, the rotation matrix is
\begin{align}
\mat{U}(\hat{a},\phi)_{ij} = \d_{ij} + \sin(\phi)\e_{ijk}\hat{a}_k+(1-\cos(\phi))\e_{ilk}\hat{a}_k\e_{ljm}\hat{a}_m
\,,
\end{align}
according to Rodrigues formula.
In order to align $\hat{n}_0\|\hat{y}$, one needs to rotate around rotation axis $\vec{a}_0=\hat{y}\times \hat{n}_0$ by an angle $\cos(\phi_0)=(\hat{y}\cdot \hat{n}_0)$ (with $\sin(\phi_0)=\abs{{a}_0}$) so that $\hat{a}_0=\vec{a}_0/\sin(\phi_0)$.
Then, in a second step one must rotate around axis $\vec{a}_1 = \hat{z}\times (\mat{U}_0\cdot\hat{b})$ with angle $\cos(\phi_1)=\hat{z}\cdot (\mat{U}_0\cdot\hat{b})$, resp. $\sin(\phi_1)=\abs{\vec{a}_1 }$ using the same procedure.
For our present slip system we find \cite{Blaschke:2020MD}
\begin{align}
\mat{U} & = \mat{U}_1\cdot \mat{U}_0
=\frac{1}{\sqrt{6}}\left(\begin{array}{ccc}
-1 & 1 & 2 \\
\sqrt{2} & -\sqrt{2} & \sqrt{2} \\
\sqrt{3} & \sqrt{3} & 0
\end{array}\right)
=\left(\begin{array}{c}
\hat{v}^T\\
\hat{n}_0^T\\
\hat{b}^T
\end{array}\right)
\,,\nn\\
\mat{U}\cdot \hat{b} &= \hat{z}
\,, \qquad\qquad U\cdot \hat{n}_0 = \hat{y}
\,, \qquad\qquad U\cdot\hat{v}=\hat{x}
\,.
\end{align}
This rotation matrix is then used to rotate the tensor of second order elastic constants \eqref{eq:C2_cubic} into the dislocation reference frame, i.e.:
\begin{align}
C'_{ijkl} = U_{ii'} U_{jj'} U_{kk'} U_{ll'}C_{i'j'k'l'} 
\,.
\end{align}
Using these elastic constants to compute the stress tensor from the ansatz $\vec{u} = (0,0,u_z(x,y,t))$ for the displacement field of a pure screw dislocation, we easily verify that the only non-vanishing stress components are
$\sigma_{13}= \sigma_{31} = \frac13(c'+2c_{44})u_{z,x} + \frac{\sqrt{2}}{3}(c44-c')u_{z,y} $
 and
$\sigma_{23}= \sigma_{32} = \frac{\sqrt{2}}{3}(c44-c')u_{z,x} + \frac1{3}(c_{44}+2c')u_{z,y}$
with $c'=(c_{11}-c_{12})/2$.
Since $\sigma_{xx}=\sigma_{xy}=\sigma_{yy}=0$, the present slip system fulfills the symmetry requirements allowing us to study pure screw dislocations.
The divergence of this stress tensor straightforwardly computes to $\partial_i\sigma_{ij} = \left(0,0,A \partial_x^2u_{z} +B\partial_x\partial_y u_z + C\partial_y^2u_z\right)$
with coefficients
\begin{align}
A & = \frac13(c'+2c_{44})\,, &
B & = \frac{2\sqrt{2}}{3}(c44-c')\,, &
C & = \frac1{3}(c_{44}+2c')
\,.\label{eq:ABCfcc}
\end{align}
One may repeat this exercise for the other 11 fcc slip systems to check that indeed all of them yield the same coefficients above.

Let us check condition $\tilde{B}^2<4\tilde{C}$, resp. $B^2<4AC$ (see \eqnref{eq:betaofpanis}) for the fcc slip systems:
$A$ and $C$ are weighted averages of the two shear moduli with $A>C$ for Zener ratio $Z\coleq c_{44}/c' >1$ and $A<C$ for $Z<1$.
$B$, on the other hand, is given by the difference of the two shear moduli and is positive for $Z>1$ and negative for $Z<1$.
Plugging \eqref{eq:ABCfcc} into the above condition yields
\begin{align}
4AC-B^2&=
 4c_{44}c'>0
\,,
\end{align}
which is clearly fulfilled for all Zener ratios.
Also notice that both $A>0$ and $C>0$, whereas $B$ can be positive or negative depending on the Zener ratio.
Our rescaled coefficients finally are $\tilde{B}=B/A=2\sqrt{2}\left(c_{44}-c'\right)/(c'+2c_{44})=2\sqrt{2}(Z-1)/(1+2Z)$ and $\tilde{C}=C/A=(2c'+c_{44})/(c'+2c_{44})=(2+Z)/(1+2Z)$.
In the isotropic limit, $Z\to1$ and hence $\tilde{B}\to0$ and $\tilde{C}\to1$.

\section{Useful relations}
\label{sec:buildingblocks}

We list some relations needed in the derivation of our main result in Section \ref{sec:laplace} above:
\begin{align}
 dp_\pm p_\pm &=\frac{\tau d\tau}{R^4}\left(x - {y}\frac{\tilde{B}}{2\tilde{C}} \pm i\frac{\abs{y}\frac{1}{\tilde{C}}\left(1 -  \frac{\tilde{B}^2}{4\tilde{C}}  \right)}{\sqrt{\frac{1}{\tilde{C}} \left(1 -  \frac{\tilde{B}^2}{4\tilde{C}}  \right) - \frac{R^2}{\tau^2 c_A^2\tilde{C}}}}\right)\left(\left(x - {y}\frac{\tilde{B}}{2\tilde{C}}\right) \pm i\abs{y}\sqrt{\frac{1}{\tilde{C}} \left(1 -  \frac{\tilde{B}^2}{4\tilde{C}}  \right) - \frac{R^2}{\tau^2 c_A^2\tilde{C}} }\right)
 \,, \\
 \Im(dp_+ p_+)&=\frac{-i}{2}(dp_+ p_+ - dp_- p_-)
 = \frac{\tau d\tau}{R^4}\left(x - {y}\frac{\tilde{B}}{2\tilde{C}} \right)\frac{\abs{y}}{\tilde{C}}\left[\frac{2\left(1 -  \frac{\tilde{B}^2}{4\tilde{C}}  \right)  - \frac{R^2}{\tau^2 c_A^2} }{\sqrt{\frac{1}{\tilde{C}} \left(1 -  \frac{\tilde{B}^2}{4\tilde{C}}  \right) - \frac{R^2}{\tau^2 c_A^2\tilde{C}}}}
 \right]
 \,,
\end{align}
\begin{align}
dp_\pm \beta_\pm &=\frac{\tau d\tau}{R^4}\!\left[ \!\left(  \frac{\abs{y}}{\tilde{C}}  - x\frac{\sgn{y}\tilde{B}}{2\tilde{C}}\right) \mp ix\sqrt{\frac{1}{\tilde{C}} \left(1 -  \frac{\tilde{B}^2}{4\tilde{C}}  \right) - \frac{R^2}{\tau^2c_A^2\tilde{C}} }\right]\!\!\left(x - {y}\frac{\tilde{B}}{2\tilde{C}} \pm i\frac{\frac{\abs{y}}{\tilde{C}}\left(1 -  \frac{\tilde{B}^2}{4\tilde{C}}  \right)}{\sqrt{\frac{1}{\tilde{C}} \left(1 -  \frac{\tilde{B}^2}{4\tilde{C}}  \right) - \frac{R^2}{\tau^2c_A^2\tilde{C}}}}\right)
,\\
\Im(dp_+ \beta_+) &=\frac{-i}{2}(dp_+ \beta_+ - dp_- \beta_-)
=\frac{\tau d\tau}{R^4}\left[ \frac{\frac{1}{\tilde{C}}\left(  \frac{y^2}{\tilde{C}}  - x\frac{y\tilde{B}}{2\tilde{C}}\right)\left(1 -  \frac{\tilde{B}^2}{4\tilde{C}}  \right) - x\left(x - {y}\frac{\tilde{B}}{2\tilde{C}} \right)\left(\frac{1}{\tilde{C}} \left(1 -  \frac{\tilde{B}^2}{4\tilde{C}}  \right) - \frac{R^2}{\tau^2c_A^2\tilde{C}} \right)}{\sqrt{\frac{1}{\tilde{C}} \left(1 -  \frac{\tilde{B}^2}{4\tilde{C}}  \right) - \frac{R^2}{\tau^2c_A^2\tilde{C}}}}
\right]
\nn\\
&=\frac{\tau d\tau}{R^4\tilde{C}}\left[ \frac{\left(  \frac{y^2}{\tilde{C}}  -x^2\right)\left(1 -  \frac{\tilde{B}^2}{4\tilde{C}}  \right) + x\left(x - {y}\frac{\tilde{B}}{2\tilde{C}} \right)\frac{R^2}{\tau^2c_A^2} }{\sqrt{\frac{1}{\tilde{C}} \left(1 -  \frac{\tilde{B}^2}{4\tilde{C}}  \right) - \frac{R^2}{\tau^2c_A^2\tilde{C}}}}\right]
\,,
\end{align}
\begin{align}
 \beta_+ p_- &= \frac{1}{\abs{y}}\left( \tau p_- - \abs{p}^2 x\right)
 \,,\\
 \Im(dp_+ \beta_+ p_-) &= \frac{1}{\abs{y}}\left( \tau \Im(dp_+ p_-) - \abs{p}^2 x \Im(dp_+ )\right)
 \nn\\
  &= \frac{d\tau}{R^2\tilde{C}\sqrt{\frac{1}{\tilde{C}} \left(1 -  \frac{\tilde{B}^2}{4\tilde{C}}  \right) - \frac{R^2}{\tau^2c_A^2\tilde{C}}}}
  \left[\frac{1}{c_A^2}\left(x - {y}\frac{\tilde{B}}{2\tilde{C}} \right) - \frac{x}{R^2}\left(\tau^2  - \frac{y^2}{c_A^2\tilde{C}} \right)\left(1 -  \frac{\tilde{B}^2}{4\tilde{C}}  \right)\right]
  \,,
\end{align}
and the primed counterparts of these expressions follow from $x\to(x-x')$.
We also need
\begin{align}
\abs{p}^2 &= p_+p_-
=\frac{1}{R^2}\left(\tau^2  - \frac{y^2}{c_A^2\tilde{C}} \right)
\,,\nn\\
\abs{\eta'(x)-p}^2 &= \frac{1}{R^2}\left(\tau^2  - \frac{y^2}{c_A^2\tilde{C}} - {2\tau}\eta'(x)\left( x - {y}\frac{\tilde{B}}{2\tilde{C}} \right) + (\eta'(x))^2{R^2}\right)
\,. \label{eq:p-abs-gen}
\end{align}
In the isotropic limit, the following simplifications apply because of $\tilde{B}\to0$, $\tilde{C}\to1$, $c_A\to\ct$:
\begin{align}
p_\pm &\to \frac{1}{r^2}\left(x\tau\pm i\abs{y}\sqrt{\tau^2 - r^2/\ct^2}\right)
\,, & 
\beta_\pm &\to \frac{1}{r^2}\left(\abs{y}\tau\mp ix\sqrt{\tau^2 - r^2/\ct^2}\right)
\,,\nn\\
dp_\pm &\to \frac{\pm i\beta_\pm d\tau}{\sqrt{\tau^2 - r^2/\ct^2}}
\,, & R^2&\to r^2 = x^2+y^2
\,.
\end{align}

\bibliographystyle{utphys-custom}
\bibliography{dislocations}

\providecommand{\accepted}[1]{accepted for publication in \textit{#1}}
\providecommand{\href}[2]{#2}\begingroup\begin{thebibliography}{10}
\small\itemsep=3pt
\tolerance 1414
\hbadness 1414
\emergencystretch 1.5em
\hfuzz 0.3pt
\widowpenalty=10000
\vfuzz \hfuzz
\raggedbottom

\bibitem{Hansen:2013}
B.~L. Hansen, I.~J. Beyerlein, C.~A. Bronkhorst, E.~K. Cerreta, and
  D.~Dennis-Koller, ``A dislocation-based multi-rate single crystal plasticity
  model'', \href{https://dx.doi.org/10.1016/j.ijplas.2012.12.006}{\emph{Int. J.
  Plast.} \textbf{44} (2013) 129--146}.

\bibitem{Luscher:2016}
D.~J. Luscher, J.~R. Mayeur, H.~M. Mourad, A.~Hunter, and M.~A. Kenamond,
  ``Coupling continuum dislocation transport with crystal plasticity for
  application to shock loading conditions'',
  \href{https://dx.doi.org/10.1016/j.ijplas.2015.07.007}{\emph{Int. J. Plast.}
  \textbf{76} (2016) 111--129}.

\bibitem{Nadgornyi:1988}
E.~M. Nadgornyi, ``Dislocation dynamics and mechanical properties of
  crystals'',
  \href{https://dx.doi.org/10.1016/0079-6425(88)90005-9}{\emph{Prog. Mater.
  Sci.} \textbf{31} (1988) 1--530}.

\bibitem{Alshits:1992}
V.~I. Alshits,
  \href{https://dx.doi.org/10.1016/B978-0-444-88773-3.50018-2}{``The
  phonon-dislocation interaction and its role in dislocation dragging and
  thermal resistivity'',} in \emph{Elastic Strain Fields and Dislocation
  Mobility}, V.~L. Indenbom and J.~Lothe, eds., vol.~31 of \emph{Modern
  Problems in Condensed Matter Sciences}, pp.~625--697, (Elsevier, 1992).

\bibitem{Blaschke:2019Bpap}
D.~N. Blaschke, E.~Mottola, and D.~L. Preston, ``Dislocation drag from phonon
  wind in an isotropic crystal at large velocities'',
  \href{https://dx.doi.org/10.1080/14786435.2019.1696484}{\emph{Phil. Mag.}
  \textbf{100} (2020) 571--600},
  \href{https://arxiv.org/abs/1907.00101}{\texttt{arXiv:1907.00101
  [cond-mat.mtrl-sci]}}.

\bibitem{Blaschke:2018anis}
D.~N. Blaschke, ``Velocity dependent dislocation drag from phonon wind and
  crystal geometry'',
  \href{https://dx.doi.org/10.1016/j.jpcs.2018.08.032}{\emph{J. Phys. Chem.
  Solids} \textbf{124} (2019) 24--35},
  \href{https://arxiv.org/abs/1804.01586}{\texttt{arXiv:1804.01586
  [cond-mat.mtrl-sci]}}.

\bibitem{Zbib:1998}
H.~M. Zbib, M.~Rhee, and J.~P. Hirth, ``On plastic deformation and the dynamics
  of {3D} dislocations'',
  \href{https://dx.doi.org/10.1016/S0020-7403(97)00043-X}{\emph{Int. J. Mech.
  Sci.} \textbf{40} (1998) 113--127}.

\bibitem{Ghoniem:2000}
N.~M. Ghoniem, S.-H. Tong, and L.~Z. Sun, ``Parametric dislocation dynamics: A
  thermodynamics-based approach to investigations of mesoscopic plastic
  deformation'', \href{https://dx.doi.org/10.1103/PhysRevB.61.913}{\emph{Phys.
  Rev.} \textbf{B61} (2000) 913--927}.

\bibitem{Bertin:2015}
N.~Bertin, M.~V. Upadhyay, C.~Pradalier, and L.~Capolungo, ``A {FFT}-based
  formulation for efficient mechanical fields computation in isotropic and
  anisotropic periodic discrete dislocation dynamics'',
  \href{https://dx.doi.org/10.1088/0965-0393/23/6/065009}{\emph{Mod. Sim.
  Mater. Sci. Eng.} \textbf{23} (2015) 065009}.

\bibitem{Cui:2019}
Y.~Cui, G.~Po, Y.-P. Pellegrini, M.~Lazar, and N.~Ghoniem, ``Computational
  3-dimensional dislocation elastodynamics'',
  \href{https://dx.doi.org/10.1016/j.jmps.2019.02.008}{\emph{J. Mech. Phys.
  Solids} \textbf{126} (2019) 20--51}.

\bibitem{Lloyd:2014JMPS}
J.~T. Lloyd, J.~D. Clayton, R.~A. Austin, and D.~L. McDowell, ``Plane wave
  simulation of elastic-viscoplastic single crystals'',
  \href{https://dx.doi.org/10.1016/j.jmps.2014.04.009}{\emph{J. Mech. Phys.
  Solids} \textbf{69} (2014) 14--32}.

\bibitem{Blaschke:2019a}
D.~N. Blaschke, A.~Hunter, and D.~L. Preston, ``Analytic model of the
  remobilization of pinned glide dislocations: including dislocation drag from
  phonon wind'',
  \href{https://dx.doi.org/10.1016/j.ijplas.2020.102750}{\emph{Int. J. Plast.}
  \textbf{131} (2020) 102750},
  \href{https://arxiv.org/abs/1912.08851}{\texttt{arXiv:1912.08851
  [cond-mat.mtrl-sci]}}.

\bibitem{Weertman:1980}
J.~Weertman and J.~R. Weertman, ``Moving dislocations'', in \emph{Moving
  Dislocations}, F.~R.~N. Nabarro, ed., vol.~3 of \emph{Dislocations in
  Solids}, pp.~1--59, (Amsterdam: North Holland Pub. Co., 1980).

\bibitem{Nosenko:2007}
V.~Nosenko, S.~Zhdanov, and G.~Morfill, ``Supersonic dislocations observed in a
  plasma crystal'',
  \href{https://dx.doi.org/10.1103/PhysRevLett.99.025002}{\emph{Phys. Rev.
  Lett.} \textbf{99} (2007) 025002},
  \href{https://arxiv.org/abs/0709.1782}{\texttt{arXiv:0709.1782
  [cond-mat.soft]}}.

\bibitem{Olmsted:2005}
D.~L. Olmsted, L.~G. Hector~Jr., W.~A. Curtin, and R.~J. Clifton, ``Atomistic
  simulations of dislocation mobility in {Al, Ni} and {Al/Mg} alloys'',
  \href{https://dx.doi.org/10.1088/0965-0393/13/3/007}{\emph{Mod. Simul. Mater.
  Sci. Eng.} \textbf{13} (2005) 371},
  \href{https://arxiv.org/abs/cond-mat/0412324}{\texttt{arXiv:cond-mat/0412324}}.

\bibitem{Marian:2006}
J.~Marian and A.~Caro, ``Moving dislocations in disordered alloys: {Connecting}
  continuum and discrete models with atomistic simulations'',
  \href{https://dx.doi.org/10.1103/PhysRevB.74.024113}{\emph{Phys. Rev.}
  \textbf{B74} (2006) 024113}.

\bibitem{Tsuzuki:2008}
H.~Tsuzuki, P.~S. Branicio, and J.~P. Rino, ``Accelerating dislocations to
  transonic and supersonic speeds in anisotropic metals'',
  \href{https://dx.doi.org/10.1063/1.2921786}{\emph{Appl. Phys. Lett.}
  \textbf{92} (2008) 191909}.

\bibitem{Oren:2017}
E.~Oren, E.~Yahel, and G.~Makov, ``Dislocation kinematics: a molecular dynamics
  study in {Cu}'',
  \href{https://dx.doi.org/10.1088/1361-651X/aa52a7}{\emph{Mod. Simul. Mater.
  Sci. Eng.} \textbf{25} (2017) 025002}.

\bibitem{Peng:2019}
S.~Peng, Y.~Wei, Z.~Jin, and W.~Yang, ``Supersonic screw dislocations gliding
  at the shear wave speed'',
  \href{https://dx.doi.org/10.1103/PhysRevLett.122.045501}{\emph{Phys. Rev.
  Lett.} \textbf{122} (2019) 045501}.

\bibitem{Blaschke:2020MD}
D.~N. Blaschke, J.~Chen, S.~Fensin, and B.~Szajewski, ``Clarifying the
  definition of `transonic' screw dislocations'',
  \href{https://dx.doi.org/10.1080/14786435.2021.1876269}{\emph{Phil. Mag.}
  \textbf{101} (2021) in press},
  \href{https://arxiv.org/abs/2008.13760}{\texttt{arXiv:2008.13760
  [cond-mat.mtrl-sci]}}.

\bibitem{Rosakis:2001}
P.~Rosakis, ``Supersonic dislocation kinetics from an augmented {Peierls}
  model'', \href{https://dx.doi.org/10.1103/PhysRevLett.86.95}{\emph{Phys. Rev.
  Lett.} \textbf{86} (2001) 95--98}.

\bibitem{Markenscoff:2008}
X.~Markenscoff and S.~Huang, ``Analysis for a screw dislocation accelerating
  through the shear-wave speed barrier'',
  \href{https://dx.doi.org/10.1016/j.jmps.2008.01.005}{\emph{J. Mech. Phys.
  Solids} \textbf{56} (2008) 2225--2239}.

\bibitem{Markenscoff:2009}
X.~Markenscoff and S.~Huang, ``The energetics of dislocations accelerating and
  decelerating through the shear-wave speed barrier'',
  \href{https://dx.doi.org/10.1063/1.3072351}{\emph{Appl. Phys. Lett.}
  \textbf{94} (2009) 021906}.

\bibitem{Huang:2009}
S.~Huang and X.~Markenscoff, ``Is intersonic dislocation motion possible?
  {Singularity} analysis for an edge dislocation accelerating through the shear
  wave speed barrier'',
  \href{https://dx.doi.org/10.1007/s11340-008-9122-8}{\emph{Exp. Mech.}
  \textbf{49} (2009) 219--224}.

\bibitem{Pillon:2007}
L.~Pillon, C.~Denoual, and Y.-P. Pellegrini, ``Equation of motion for
  dislocations with inertial effects'',
  \href{https://dx.doi.org/10.1103/PhysRevB.76.224105}{\emph{Phys. Rev.}
  \textbf{B76} (2007) 224105},
  \href{https://arxiv.org/abs/0707.0645}{\texttt{arXiv:0707.0645
  [cond-mat.mtrl-sci]}}.

\bibitem{Pellegrini:2010}
Y.-P. Pellegrini, ``Dynamic {Peierls-Nabarro} equations for elastically
  isotropic crystals'',
  \href{https://dx.doi.org/10.1103/PhysRevB.81.024101}{\emph{Phys. Rev.}
  \textbf{B81} (2010) 024101},
  \href{https://arxiv.org/abs/0908.2371}{\texttt{arXiv:0908.2371
  [cond-mat.mtrl-sci]}}.

\bibitem{Pellegrini:2014}
Y.-P. Pellegrini, ``Equation of motion and subsonic-transonic transitions of
  rectilinear edge dislocations: {A} collective-variable approach'',
  \href{https://dx.doi.org/10.1103/PhysRevB.90.054120}{\emph{Phys. Rev.}
  \textbf{B90} (2014) 054120},
  \href{https://arxiv.org/abs/1307.5244}{\texttt{arXiv:1307.5244
  [cond-mat.mtrl-sci]}}.

\bibitem{Pellegrini:2020}
Y.-P. Pellegrini, ``Dynamic {Peach-Koehler} self-force, inertia, and radiation
  damping of a regularized dislocation'',
  \href{https://arxiv.org/abs/2005.12704}{\texttt{arXiv:2005.12704
  [cond-mat.mtrl-sci]}}.

\bibitem{Bacon:1980}
D.~J. Bacon, D.~M. Barnett, and R.~O. Scattergood, ``Anisotropic continuum
  theory of lattice defects'',
  \href{https://dx.doi.org/10.1016/0079-6425(80)90007-9}{\emph{Prog. Mater.
  Sci.} \textbf{23} (1980) 51--262}.

\bibitem{Pellegrini:2017}
Y.-P. Pellegrini, ``{Causal Stroh formalism for uniformly-moving dislocations
  in anisotropic media: Somigliana dislocations and Mach cones}'',
  \href{https://dx.doi.org/10.1016/j.wavemoti.2016.09.006}{\emph{Wave Motion}
  \textbf{68} (2017) 128--148},
  \href{https://arxiv.org/abs/1609.02749}{\texttt{arXiv:1609.02749
  [cond-mat.mtrl-sci]}}.

\bibitem{Pellegrini:2018}
Y.-P. Pellegrini, ``Uniformly-moving non-singular dislocations with ellipsoidal
  core shape in anisotropic media'',
  \href{https://dx.doi.org/10.1142/S2424913018400040}{\emph{J. Micromech.
  Molec. Phys.} \textbf{3} (2018) 1840004},
  \href{https://arxiv.org/abs/1808.10272}{\texttt{arXiv:1808.10272
  [physics.class-ph]}}.

\bibitem{Markenscoff:1984JE}
X.~Markenscoff and L.~Ni, ``The transient motion of a screw dislocation in an
  anisotropic medium'', \href{https://dx.doi.org/10.1007/BF00041084}{\emph{J.
  Elast.} \textbf{14} (1984) 93--95}.

\bibitem{Markenscoff:1984}
X.~Markenscoff and L.~Q. Ni, ``Nonuniform motion of an edge dislocation in an
  anisotropic solid. {I}'',
  \href{https://dx.doi.org/10.1090/qam/724058}{\emph{Quart. Appl. Math.}
  \textbf{41} (1984) 475--494}.

\bibitem{Markenscoff:1985a}
X.~Markenscoff and L.~Ni, ``Nonuniform motion of an edge dislocation in an
  anisotropic solid. {II}'',
  \href{https://dx.doi.org/10.1090/qam/766879}{\emph{Quart. Appl. Math.}
  \textbf{42} (1985) 425--432}.

\bibitem{Payton:1985}
R.~G. Payton, ``Transient stresses in a transversely isotropic elastic solid
  caused by a moving dislocation'',
  \href{https://dx.doi.org/10.1007/BF00945456}{\emph{Z. Angew. Math. Phys.}
  \textbf{36} (1985) 191--203}.

\bibitem{Payton:1995}
R.~G. Payton, ``Steady state stresses induced in a transversely isotropic
  elastic solid by a moving dislocation'',
  \href{https://dx.doi.org/10.1007/BF00944758}{\emph{Z. Angew. Math. Phys.}
  \textbf{46} (1995) 282--288}.

\bibitem{Gurrutxaga:2020}
B.~Gurrutxaga-Lerma, J.~Verschueren, A.~P. Sutton, and D.~Dini, ``The mechanics
  and physics of high-speed dislocations: a critical review'',
  \href{https://dx.doi.org/10.1080/09506608.2020.1749781}{\emph{Int. Mater.
  Rev.} (2020) in press}.

\bibitem{Hirth:1982}
J.~P. Hirth and J.~Lothe, \emph{Theory of Dislocations}, second~ed., (New York:
  Wiley, 1982).

\bibitem{Teutonico:1961}
L.~J. Teutonico, ``Dynamical behavior of dislocations in anisotropic media'',
  \href{https://dx.doi.org/10.1103/PhysRev.124.1039}{\emph{Phys. Rev.}
  \textbf{124} (1961) 1039--1045}.

\bibitem{Cagniard:1939}
L.~Cagniard,
  \href{http://www.numdam.org/item/THESE_1939__225__R1_0}{\emph{R{\'e}flexion
  et r{\'e}fraction des ondes s{\'e}ismiques progressives}}.
\newblock PhD thesis, Universit\'e de Paris, Sorbonne, 1939.

\bibitem{DeHoop:1960}
A.~T. De~Hoop, ``A modification of {Cagniard's} method for solving seismic
  pulse problems'', \href{https://dx.doi.org/10.1007/BF02920068}{\emph{Appl.
  Sci. Res.} \textbf{8} (1960) 349--356}.

\bibitem{Freund:1973}
L.~B. Freund, ``The response of an elastic solid to nonuniformly moving surface
  loads'', \href{https://dx.doi.org/10.1115/1.3423076}{\emph{J. Appl. Mech.}
  \textbf{40} (1973) 699--704}.

\bibitem{Markenscoff:1980}
X.~Markenscoff, ``The transient motion of a nonuniformly moving dislocation'',
  \href{https://dx.doi.org/10.1007/BF00044503}{\emph{J. Elast.} \textbf{10}
  (1980) 193--201}.

\bibitem{Mitra:1966}
M.~Mitra, ``Surface displacement produced by an underground fracture'',
  \href{https://dx.doi.org/10.1190/1.1439735}{\emph{Geophysics} \textbf{31}
  (1966) 204--213}.

\bibitem{Boore:1971}
D.~M. Boore, K.~Aki, and T.~Todd, ``{A two-dimensional moving dislocation model
  for a strike-slip fault}'',
  \href{https://pubs.geoscienceworld.org/bssa/article-pdf/61/1/177/2697155/BSSA0610010177.pdf}{\emph{Bull.
  Seismol. Soc. Am.} \textbf{61} (1971) 177--194}.

\bibitem{Boore:1974}
D.~M. Boore and M.~D. Zoback, ``{Near-field motions from kinematic models of
  propagating faults}'',
  \href{https://pubs.geoscienceworld.org/bssa/article-pdf/64/2/321/2698644/BSSA0640020321.pdf}{\emph{Bull.
  Seismol. Soc. Am.} \textbf{64} (1974) 321--342}.

\bibitem{Madariaga:1978}
R.~Madariaga, ``{The dynamic field of Haskell's rectangular dislocation fault
  model}'',
  \href{https://pubs.geoscienceworld.org/bssa/article-pdf/68/4/869/2701127/BSSA0680040869.pdf}{\emph{Bull.
  Seismol. Soc. Am.} \textbf{68} (1978) 869--887}.

\bibitem{Kuhfittig:1978}
P.~K.~F. Kuhfittig,
  \href{https://dx.doi.org/10.1007/978-1-4899-2201-4}{\emph{Introduction to the
  Laplace Transform}}, vol.~8 of \emph{Mathematical concepts and methods in
  science and engineering}, A.~Miele, ed., (Springer-Verlag, 1978).

\bibitem{Markenscoff:1985}
X.~Markenscoff, ``The singularities of nonuniformly moving dislocations'',
  \href{https://dx.doi.org/10.1016/0020-7683(85)90079-4}{\emph{Int. J. Solids
  Struct.} \textbf{21} (1985) 767--772}.

\bibitem{CRCHandbook}
J.~R. Rumble, ed., \href{http://hbcponline.com}{\emph{CRC Handbook of Chemistry
  and Physics}}, 100th~ed., (CRC Press, 2019).

\bibitem{Blaschke:2019fits}
D.~N. Blaschke, ``Properties of dislocation drag from phonon wind at ambient
  conditions'', \href{https://dx.doi.org/10.3390/ma12060948}{\emph{Materials}
  \textbf{12} (2019) 948},
  \href{https://arxiv.org/abs/1902.02451}{\texttt{arXiv:1902.02451
  [cond-mat.mtrl-sci]}}.

\bibitem{Callias:1980}
C.~Callias and X.~Markenscoff, ``The nonuniform motion of a supersonic
  dislocation'', \href{https://dx.doi.org/10.1090/qam/592199}{\emph{Quart.
  Appl. Math.} \textbf{38} (1980) 323--330}.

\bibitem{Callias:1988}
C.~Callias and X.~Markenscoff, ``Singular asymptotics of integrals and the
  near-field radiated from nonuniformly moving dislocations'',
  \href{https://dx.doi.org/10.1007/BF00281350}{\emph{Arch. Ration. Mech. Anal.}
  \textbf{102} (1988) 273--285}.

\bibitem{pydislocdyn}
D.~N. Blaschke, \emph{{PyDislocDyn}}, 2018--2021, \textsc{url:}
  \url{https://github.com/dblaschke-LANL/PyDislocDyn}.

\bibitem{Blaschke:2021impact}
D.~N. Blaschke and D.~J. Luscher, ``Dislocation drag and its influence on
  elastic precursor decay'',
  \href{https://arxiv.org/abs/2101.10497}{\texttt{arXiv:2101.10497
  [cond-mat.mtrl-sci]}}.

\bibitem{Austin:2018}
R.~A. Austin, ``Elastic precursor wave decay in shock-compressed aluminum over
  a wide range of temperature'',
  \href{https://dx.doi.org/10.1063/1.5008280}{\emph{J. Appl. Phys.}
  \textbf{123} (2018) 035103}.

\bibitem{Eshelby:1949}
J.~D. Eshelby, ``Uniformly moving dislocations'',
  \href{https://dx.doi.org/10.1088/0370-1298/62/5/307}{\emph{Proc. Phys. Soc.}
  \textbf{A62} (1949) 307}.

\bibitem{Markenscoff:2001}
X.~Markenscoff and L.~Ni, ``The transient motion of a dislocation with a
  ramp-like core'',
  \href{https://dx.doi.org/10.1016/S0022-5096(00)00062-4}{\emph{J. Mech. Phys.
  Solids} \textbf{49} (2001) 1603--1619}.

\bibitem{Clouet:2011a}
E.~Clouet, ``Dislocation core field. {I. Modeling} in anisotropic linear
  elasticity theory'',
  \href{https://dx.doi.org/10.1103/PhysRevB.84.224111}{\emph{Phys. Rev.}
  \textbf{B84} (2011) 224111},
  \href{https://arxiv.org/abs/1112.4938}{\texttt{arXiv:1112.4938
  [cond-mat.mtrl-sci]}}.

\bibitem{Szajewski:2017phil}
B.~A. Szajewski, A.~Hunter, and I.~J. Beyerlein, ``The core structure and
  recombination energy of a copper screw dislocation: a {Peierls} study'',
  \href{https://dx.doi.org/10.1080/14786435.2017.1328138}{\emph{Phil. Mag.}
  \textbf{97} (2017) 2143--2163}.

\bibitem{Boleininger:2019}
M.~Boleininger and S.~L. Dudarev, ``Continuum model for the core of a straight
  mixed dislocation'',
  \href{https://dx.doi.org/10.1103/PhysRevMaterials.3.093801}{\emph{Phys. Rev.
  Mater.} \textbf{3} (2019) 093801}.

\bibitem{Gurrutxaga:2019}
B.~Gurrutxaga-Lerma and J.~Verschueren, ``Generalized {Kanzaki} force field of
  extended defects in crystals with applications to the modeling of edge
  dislocations'',
  \href{https://dx.doi.org/10.1103/PhysRevMaterials.3.113801}{\emph{Phys. Rev.
  Mater.} \textbf{3} (2019) 113801}.

\end{thebibliography}\endgroup

\end{document}